\documentstyle[12pt,aps]{revtex}

\begin{document}
\draft
\title{QED in external field with space-time uniform invariants. Exact solutions }
\author{S.P. Gavrilov\thanks{%
On leave from Tomsk Pedagogical University, 634041 Tomsk, Russia; present
e-mail: gavrilov@sergipe.ufs.br}, D.M. Gitman\thanks{%
Instituto de F\'\i sica, Universidade de S\~ao Paulo, Caixa Postal 66318,
05315-970-S\~ao Paulo, SP, Brazil; e-mail: gitman@fma1.if.usp.br} and A.E.
Gon\c calves \thanks{%
Dept. F\'\i sica, CCE, Universidade Estadual de Londrina, CP 6001. CEP
86051-990, Londrina, PR, Brasil; e-mail: goncalve@fisica.uel.br}}
\address{Dept. de F\'\i sica, CCET, Universidade Federal de Sergipe,\\
49000-000 Aracaju, SE, Brazil}
\date{\today}
\maketitle

\begin{abstract}
We study exact solutions of Dirac and Klein-Gordon equations and Green
functions in $d$-dimensional QED and in an external electromagnetic field
with constant and homogeneous field invariants. The cases of even and odd
dimensions are considered separately, they are essentially different. We
direct attention to the asymmetry of the quasienergy spectrum, which appears
in odd dimensions. The $in$ and $out$ classification of the exact solutions
as well as the completeness and orthogonality relations is strictly proven.
Different Green functions in the form of sums over the exact solutions are
constructed. The Fock-Schwinger proper time integral representations of
these Green functions are found. As physical applications we consider the
calculations of different quantum effects related to the vacuum instability
in the external field. For example, we present mean values of particles
created from the vacuum, the probability of the vacuum remaining a vacuum,
the effective action, and the expectation values of the current and
energy-momentum tensor.
\end{abstract}

\pacs{11.10.Kk, 04.62.+v, 12.20.Ds}

\section{Introduction}

Exact solutions of the relativistic wave equations (Klein-Gordon or Dirac
equation) in an external electromagnetic field as well as Green functions of
those equations are very important in QED with such a field. Special
complete sets of the exact solutions and special kinds of the Green
functions allow one to calculate different quantum effects, for example,
particles scattering, pair creation and so on, in zero order with respect to
the radiative corrections, but taking the interaction with the external
field into account exactly \cite{GMR,GMM,FGS}. They may serve also as a
basis for the perturbation theory with respect to the radiative interaction,
in which the external field is taken into account exactly (so called
generalized Furry picture) \cite{F,GG1977}. In four-dimensional space-time
the above exact solutions were studied for different configurations of the
external field in numerous papers and books (see, for example, \cite{FGS,BG}%
. Among all the configurations of the external electromagnetic fields, which
admit the exact solutions in $d=4$, one is especially important due to the
fact that it corresponds to a wide class of real physical situations. It is
a combination of constant uniform electric and magnetic fields and plane
wave field. The exact solutions can be found if there exist a reference
frame in which the constant electric and magnetic fields, and the wave
3-vector are collinear. First, the exact solutions of Klein-Gordon and Dirac
equation in such a configuration were found in \cite{BGJ}. Some of the Green
functions of scalar field in the same configuration were studied in the
papers \cite{GGS}. Using those exact solutions and Green functions different
quantum effects were calculated \cite{GGSVN}. Calculation of Green functions
of spinor field was not present in the literature. Moreover, lately, field
theoretical models in dimensions different from $d=4$ attract attention due
to various reasons. One can mention here models in $2+1$ dimensions which
probably describe planar physical phenomenon, and models in $d>4$ dimensions
in connection with Kaluza-Klein ideas (see e.g. \cite{BOS}. That is why in
the present article we are going to study exact solutions and Green
functions (both scalar and spinor case) in $d$-dimensional QED with an
external electromagnetic field, which may be considered as a generalization
of the above mentioned special configuration in four dimensions. At the same
time we also present new results in four dimensions. A general definition of
such a configuration valid for any dimensions of space-time is the
following: an external electromagnetic field with constant and homogeneous
field invariants. Also, an important motivation to consider exact solutions
and Green functions in such a configuration of the external electromagnetic
field is the fact that formally problems in some configurations of an
external gravitational field lead to the same kind of equations, so that the
former exact solutions can be used in the latter case after some simple
identifications. Such a reduction may be done, for example, for De Sitter
and FRW metrics \cite{SD,GGO}, for four fermionic models in an external
field and so on.

In a sense, the paper can be considered as a continuation and generalization
of the one \cite{GG1996a}, where only an electric-like external field was
considered. The latter paper can help the reader to better understand the
logic of the exact solutions constructed in this more complicated present
case. Thus, here we sometimes omit technical details and explanations which
can be extracted from the above mentioned previous paper.

The paper is organized as follows: In Sec. II we describe first the external
electromagnetic field under consideration. Then, we present the so called in
and out complete sets of exact solutions of Dirac and Klein-Gordon equations
in the external field. The cases of even and odd dimensions have to be
considered separately, they are essentially different. We attract attention
to the asymmetry of the quasienergy spectrum, which appears in odd
dimensions. We spend enough time to prove strictly the $in$ and $out$
classification of the exact solutions as well as to prove the completeness
and orthogonality relations. Then in Sec. III we construct different Green
functions in the form of sums over the exact solutions and present them in
the form of contour integrals over the Fock-Schwinger proper time. Among
them are the causal Green function, $in-in$ and $out-out$ Green functions,
the commutator function and so on. One ought to say that the spinor case is
treated first, even in $d=4$ dimensions. As physical applications we
consider in Sec. IV the calculations of different quantum effects related to
the vacuum instability in the external field. For example, we present mean
values of particles created from the vacuum, the probability for the vacuum
remaining a vacuum, the effective action, the expectation values of the
current and energy-momentum tensor. We calculate the latter quantities by
means of both $in$ and $out$ sets of the exact solutions.

\section{Complete sets of exact solutions}

The Dirac equation in an external electromagnetic field with potentials $%
A_\mu (x)$ in $d$ dimensions has the form ($\hbar =c=1$) 
\begin{equation}
\left( {\cal P}_\mu \gamma ^\mu -m\right) \psi (x)=0\;,\;\;\;{\cal P}_\mu
=i\partial _\mu -qA_\mu (x)\;,  \label{e1}
\end{equation}
where $\psi (x)$ is a $2^{[\frac d2]}$-component column, $\gamma ^\mu $ are $%
\gamma $-matrices in $d$ dimensions \cite{BrW}, 
\[
\lbrack \gamma ^\mu ,\gamma ^\nu ]_{+}=2\eta ^{\mu \nu },\;\;\eta ^{\mu \nu
}={\rm diag}(\underbrace{1,-1,-1,\ldots }_d),\;\;d=D+1, 
\]
and $x=(x^\mu )=(x^0,{\bf x}),\;{\bf x}=(x^i),\;\;\mu =0,1,\ldots
,D,\;\;i=1,\ldots ,D$. The time-independent scalar product of the solutions
of the Dirac equation may be chosen in the conventional form 
\begin{equation}
(\psi ,\psi ^{\prime })=\int \bar \psi (x)\gamma ^0\psi ^{\prime }(x)d{\bf x}%
.  \label{w1}
\end{equation}

As usual, it is convenient to present $\psi (x)$ in the form 
\begin{equation}
\psi (x)=\left( {\cal P}_\mu \gamma ^\mu +m\right) \phi (x)\;.  \label{e2}
\end{equation}
Then the functions $\phi $ have to obey the squared Dirac equation in $d$
dimensions, 
\begin{equation}
\left( {\cal P}^2-m^2-\frac q2\sigma ^{\mu \nu }{\cal F}_{\mu \nu }\right)
\phi (x)=0\;,\;\;\;{\cal F}_{\mu \nu }=\partial _\mu A_\nu (x)-\partial _\nu
A_\mu (x)\;,\;\sigma ^{\mu \nu }=\frac i2[\gamma ^\mu ,\gamma ^\nu ]\;.
\label{e3}
\end{equation}

To construct the above mentioned generalized Furry picture in QED with an
external field one has to find special sets of exact solutions of Eq. (\ref
{e1}), namely, two complete and orthonormal sets of solution: $\left\{ _{\pm
}\psi _{\left\{ n\right\} }(x)\right\} $ which describes particles (+) and
antiparticles ($-$) in the initial time instant ($x^{0}\rightarrow -\infty )$%
, and $\left\{ ^{\pm }\psi _{\left\{ n\right\} }(x)\right\} $ which
describes particles (+) and antiparticles ($-$) in the final time instant ($%
x^{0}\rightarrow +\infty ).$ According to the general approach \cite{GG1977}
such solutions obey the following asymptotic conditions 
\begin{eqnarray}
\ &&H_{o.p.}(x^{0})\;{}_{\zeta }\psi _{\left\{ n\right\} }(x)={}_{\zeta
}\varepsilon \;{}_{\zeta }\psi _{\left\{ n\right\} }(x),\;\;\;,{\rm sgn}%
\;{}_{\zeta }\varepsilon =\zeta ,\;x^{0}\rightarrow -\infty \;,  \nonumber \\
\ &&H_{o.p.}(x^{0})\;{}^{\zeta }\psi _{\left\{ l\right\} }(x)={}^{\zeta
}\varepsilon \;{}^{\zeta }\psi _{\left\{ l\right\} }(x),\;{\rm sgn}%
\;{}^{\zeta }\varepsilon =\zeta ,\;\;x^{0}\rightarrow +\infty \;,  \label{w2}
\end{eqnarray}
where $\zeta ,\left\{ n\right\} $ and $\zeta ,\left\{ l\right\} $ are
complete sets of quantum numbers which characterize solutions $_{\zeta }\psi
_{\left\{ n\right\} }(x)$ and $^{\zeta }\psi _{\left\{ l\right\} }(x)$
respectively, $H_{o.p.}=\gamma ^{0}(m-\gamma ^{i}{\cal P}_{i})$ is a
one-particle Dirac Hamiltonian in convenient external field gauge $%
A_{0}(x)=0 $; $^{+}\varepsilon $, $_{+}\varepsilon $ are particle
quasi-energies and $|^{-}\varepsilon |$ and $|_{-}\varepsilon |$ are
antiparticles quasi-energies. All the information about the processes of
particles scattering and creation by an external field (in zeroth order with
respect to the radiative corrections) can be extracted from the
decomposition coefficients (matrices) $G\left( {}_{\zeta }|{}^{\zeta
^{\prime }}\right) $, 
\begin{equation}
{}^{\zeta }\psi (x)={}_{+}\psi (x)G\left( {}_{+}|{}^{\zeta }\right)
+{}_{-}\psi (x)G\left( {}_{-}|{}^{\zeta }\right) \;.  \label{we10}
\end{equation}
The matrices $G\left( {}_{\zeta }|{}^{\zeta ^{\prime }}\right) $ obey the
following relations, 
\begin{eqnarray}
\ &&G\left( {}_{\zeta }|{}^{+}\right) G\left( {}_{\zeta }|{}^{+}\right)
^{\dagger }+G\left( {}_{\zeta }|{}^{-}\right) G\left( {}_{\zeta
}|{}^{-}\right) ^{\dagger }={\bf I},  \nonumber \\
\ &&G\left( {}_{+}|{}^{+}\right) G\left( {}_{-}|{}^{+}\right) ^{\dagger
}+G\left( {}_{+}|{}^{-}\right) G\left( {}_{-}|{}^{-}\right) ^{\dagger }=0\;,
\label{w15}
\end{eqnarray}
where ${\bf I}$ is the identity matrix.

Let us describe the external electromagnetic field in which we are going to
construct the exact solutions. Such a field is a generalization of the
corresponding field in $d=4$, which is a combination of the constant uniform
field with the plane wave field and which admits there exact solutions. One
can describe such a field in $d=4$ in arbitrary reference frame saying that
both its field invariants do not depend on space-time coordinates. In $d>4$
there exist, generally speaking, more independent field invariants. Besides $%
I_{1}=1/2{\cal F}_{\mu \nu }{\cal F}^{\mu \nu }$, those are all possible
independent invariant combinations which may be constructed from the field
tensor ${\cal F}_{\mu \nu }$ and Levi- Civita tensor. It is easy to see that
there exist $\left[ d/2\right] $ such invariants. We define the external
field under consideration in arbitrary dimensions in the same manner, all
its field invariants have to be constant and uniform. One can see that such
a field is a combination of a constant uniform field $F_{\mu \nu }$ and a
plane-wave field $f_{\mu \nu }(nx)$ 
\begin{equation}
{\cal F}_{\mu \nu }=F_{\mu \nu }+f_{\mu \nu }(nx),  \label{w5}
\end{equation}
where $n_{\mu }$ is an real isotropic vector, $n_{\mu }n^{\mu }=0$. It is an
eigenvector of the tensor $F_{\mu \nu },$ $F_{\mu \nu }n^{\nu }={\cal E}%
n_{\mu },\smallskip \ ,$ and $f_{\mu \nu }(nx)$ is a transverse field with
respect to $n_{\mu },$ $n^{\mu }f_{\mu \nu }(nx)=f_{\mu \nu }(nx)n^{\nu }=0.$
If all of the invariants are equal to zero (it is only possible for $d>2$),
only a plane-wave field configuration is possible. If some of the invariants
are not equal to zero, then a constant, uniform component $F_{\mu \nu },$ is
not zero. The field (\ref{w5}) is free. Such a field is of special interest
due to the fact that QED with a free classical field can be treated as exact
QED (without external fields) but with some special (coherent) initial
photon states \cite{BGK,FGS}.

If the eigenvalue ${\cal E}$ is not zero, the external field violates the
vacuum stability (creating particles). One can find an inertial frame where
the matrix ${\cal F}_{\mu \nu }$ has a simple form 
\begin{eqnarray}
&&F_{\mu \nu }=F_{\mu \nu }^{\bot }+F_{\mu \nu }^{\Vert },  \label{w6} \\
&&F_{\mu \nu }^{\Vert }=E\left( \delta _{\mu }^{0}\delta _{\nu }^{D}-\delta
_{\mu }^{D}\delta _{\nu }^{0}\right) ,  \nonumber \\
&&f_{\mu \nu }\left( nx\right) =\sum_{k=1}^{D-1}\left( n_{\mu }\delta _{\nu
}^{k}-n_{\nu }\delta _{\mu }^{k}\right) \dot{f}_{k}\left( x_{-}\right)
,\smallskip \ \dot{f}_{k}(x_{-})=\frac{df_{k}(x_{-})}{dx_{-}},  \nonumber \\
&&n_{\mu }=\delta _{\mu }^{0}-\delta _{\mu }^{D},\;\;x_{-}=nx=x^{0}-x^{D}\;.
\nonumber
\end{eqnarray}
If $d>2$ is even then 
\[
F_{\mu \nu }^{\perp }=\sum_{j=1}^{\left( d-2\right) /2}H_{j}(\delta _{\mu
}^{2j}\delta _{\nu }^{2j-1}-\delta _{\nu }^{2j}\delta _{\mu }^{2j-1})\,, 
\]
and if $d=2$ the fields $F_{\mu \nu }^{\perp }$ and $f_{\mu \nu }\left(
nx\right) $ are absent. In the case $d$ is odd, and ${\cal E}\neq 0,$ 
\[
F_{\mu \nu }^{\perp }=\sum_{j=1}^{\left( d-3\right) /2}H_{j}(\delta _{\mu
}^{2j}\delta _{\nu }^{2j-1}-\delta _{\nu }^{2j}\delta _{\mu
}^{2j-1})\smallskip \ {\rm if}\text{ }d>3,\smallskip \ {\rm and}\smallskip \
F_{\mu \nu }^{\perp }=0\smallskip \ \text{{\rm if }}d=3\,. 
\]
The same formula is valued if ${\cal E}=0$, but at least one of the
imaginary eigenvalues of $F_{\mu \nu }$ is zero. Here $f_{k}(x_{-})$ are
arbitrary functions of $x_{-}$. If ${\cal E}=0$, and all imaginary
eigenvalues of $F_{\mu \nu }$ are not equal to zero (all field invariants
are not equal to zero) we have 
\[
E=0,\smallskip \ F_{\mu \nu }^{\perp }=\sum_{j=1}^{\left( d-1\right)
/2}H_{j}(\delta _{\mu }^{2j}\delta _{\nu }^{2j-1}-\delta _{\nu }^{2j}\delta
_{\mu }^{2j-1}),\smallskip \ \smallskip \ f_{k}\left( x_{-}\right)
=0\smallskip \ {\rm for}\smallskip \ {\rm all}\smallskip \ k\,. 
\]
In the reference frame under consideration ${\cal E}=E.$

To realize the external electromagnetic field of the above form we select
the following potentials: 
\begin{eqnarray}
&&A_\mu \left( x\right) =A_\mu ^E\left( x\right) +A_\mu ^H\left( x\right)
+f_\mu \left( x_{-}\right) ,  \label{w9} \\
&&A_\mu ^E\left( x\right) =Ex^0\delta _\mu ^D,\smallskip\
A_i^H=-H_j\,x_{i+1}\,\delta _{i,2j-1},\;j=1,\ldots ,[d/2]-1,\;i=1,\ldots
,D-1,  \nonumber \\
&&f_\mu \left( x_{-}\right) =0\smallskip\ {\rm if}\smallskip\ \mu =0,D\;. 
\nonumber
\end{eqnarray}

Below we present linearly independent sets of solutions of the squared Dirac
equation, which correspond to the particles in the initial time instance and
to the antiparticles in the final time instant: 
\begin{eqnarray}
\ {}_{+}^{-}\phi _{p_{-},n,\xi ,r}(x) &=&\;{}_{+}^{-}\phi
_{p_{-},n,r}(x^{0},x^{D})\;{}_{+}^{-}u_{\xi ,r}(x_{-})\phi _{n,r}(x_{\perp
}),  \label{w10} \\
\ \phi _{n,r}(x_{\perp }) &=&\phi _{p_{1},n_{1}}(x_{1},x_{2})\phi
_{p_{3},n_{2}}(x_{3},x_{4})...\phi
_{p_{d-3},n_{(d-2)/2}}(x_{D-2},x_{D-1}),\;\;\mbox{if $d$
is even},  \nonumber \\
\ \phi _{n,r}(x_{\perp }) &=&\phi _{p_{1},n_{1}}(x_{1},x_{2})\phi
_{p_{3},n_{2}}(x_{3},x_{4})...\phi _{p_{d-2},n_{(d-3)/2}}(x_{D-3},x_{D-2}) 
\nonumber \\
\ &&\ (2\pi )^{-1/2}\exp (-ip_{D-1}x^{D-1})\;\;\mbox{if $d$ is odd }{\rm and}%
\smallskip \ H_{(d-1)/2}=0,  \nonumber \\
\ {}_{+}^{-}\phi _{p_{-},n,r}(x^{0},x^{D}) &=&(4\pi )^{-1/2}\exp \left\{ 
\frac{i}{2}\left( qE(x_{-}^{2}/2-x_{D}^{2})-p_{-}x_{+}+\lambda \ln \left(
\mp \tilde{\pi}_{-}\right) \right) \;+\right.  \nonumber \\
&&\ \ \left. {}_{+}^{-}J(x_{-})-{}_{+}^{-}K^{\mu }(x_{-})\pi _{\perp \mu
})\right\} ,  \nonumber \\
_{+}^{-}J\left( x_{-}\right) &=&-\frac{1}{2qE}\int_{\mp \infty }^{\pi
_{-}}qf\left( \frac{p_{-}-\tau }{qE}\right) \left[ qf\left( \frac{p_{-}-\tau 
}{qE}\right) +qF\;_{+}^{-}K\left( \frac{p_{-}-\tau }{qE}\right) \right] \tau
^{-1}d\tau ,  \nonumber \\
_{+}^{-}K\left( x_{-}\right) &=&-\frac{1}{qE}\int_{\mp \infty }^{\pi
_{-}}\exp \left\{ -\frac{F}{E}\ln \frac{\pi _{-}}{\tau }\right\} qf\left( 
\frac{p_{-}-\tau }{qE}\right) \tau ^{-1}d\tau ,  \nonumber \\
{}_{+}^{-}u_{\xi ,r}(x_{-})=\mp &&\left( \frac{1}{2\pi _{-}}(1+\gamma
^{0}\gamma ^{D})+\frac{1}{2}(1-\gamma ^{0}\gamma ^{D})+\frac{1}{2\pi _{-}}%
(\gamma ^{0}-\gamma ^{D})\gamma qf(x_{-})\right) v_{\xi ,r},  \nonumber \\
&&x_{+}=x^{0}+x^{D},\,\,\tilde{\pi}_{-}=\pi _{-}/\sqrt{qE},  \nonumber
\end{eqnarray}
where $p_{-},$ $n=\left( n_{1},n_{2},\ldots ,n_{[d/2]-1};p_{1},p_{3},\ldots
,p_{2[(d-1)/2]-1}\right) ,\;\xi $ and $r=(r_{1},r_{2},\ldots ,r_{[d/2]-1})$
is a complete set of quantum numbers. Among them $p_{-},$ $p_{j}$ are
momenta of the continuous spectrum, $n_{j}$ are integer quantum numbers, $%
\xi =\pm 1$ and $r_{j}=\pm 1$ are spin quantum numbers; the momentum $p_{-}$
is the eigenvalue of the operator $2i\frac{\partial }{\partial x_{+}}$; if $%
qE>0$ is chosen then the signs $\mp $ assigned to the functions $\
{}_{+}^{-}\phi _{p_{-},n,\xi ,r}(x)$ are matched with those of the kinetic
momentum $\pi _{-}=p_{-}-qEx_{-}$ at $x_{-}\rightarrow \pm \infty $, and 
\begin{eqnarray}
&&x_{\perp }^{\mu }=0\smallskip \ {\rm if}\smallskip \ \mu =0,D,\smallskip \
x_{\perp }^{\mu }=x^{\mu }\;{\rm if}\smallskip \ \mu =1,\ldots ,D-1, 
\nonumber \\
&&\pi _{\perp \mu }=0\smallskip \ {\rm if}\smallskip \ \mu =0,D,\smallskip \
\ \pi _{\perp \mu }=i\frac{\partial }{\partial x^{\mu }}-qA_{\mu }^{H}(x)\;%
{\rm if}\smallskip \ \mu =1,\ldots ,D-1,  \nonumber \\
&&\ qE\lambda =m^{2}+\sum_{j=1}^{[d/2]-1}\omega _{j}+\omega
_{0},\;\;\;\omega _{0}=\left\{ 
\begin{array}{ll}
0, & d\;\mbox{is even} \\ 
p_{d-2}^{2}, & d\;\mbox{is odd}\;,
\end{array}
\right.  \label{w11} \\
&&\omega _{j}=\left\{ 
\begin{array}{ll}
|qH_{j}|(2n_{j}+1-r_{j}),\; & H_{j}\neq 0 \\ 
p_{2j-1}^{2}+p_{2j}^{2}\;\;, & H_{j}=0
\end{array}
\right. \;.  \nonumber
\end{eqnarray}
Each function $\phi _{p_{2j-1},n_{j}}(x_{2j-1},x_{2j})$ obeys the following
equations 
\begin{eqnarray*}
(\pi _{\perp 2j-1}^{2}+\pi _{\perp 2j}^{2}-\omega _{j})\phi
_{p_{2j-1},n_{j}}(x_{2j-1},x_{2j}) &=&0, \\
\left( \pi _{\perp 2j-1}-p_{2j-1}\right) \phi
_{p_{2j-1},n_{j}}(x_{2j-1},x_{2j}) &=&0.
\end{eqnarray*}
If $H_{j}\neq 0,$ a solution of these equations is 
\begin{eqnarray*}
&&\phi _{p_{2j-1},n_{j}}(x_{2j-1},x_{2j})= \\
&&\left( \frac{\sqrt{|qH_{j}|}}{2^{n_{j}+1}\pi ^{\frac{3}{2}}n_{j}!}\right)
^{1/2}\exp \left\{ -ip_{2j-1}x^{2j-1}-\frac{|qH_{j}|}{2}\left( x^{2j}+\frac{%
p^{2j-1}}{qH_{j}}\right) ^{2}\right\} {\cal H}_{n_{j}}\left[ \sqrt{|qH_{j}|}%
\left( x^{2j}+\frac{p^{2j-1}}{qH_{j}}\right) \right] ,
\end{eqnarray*}
where ${\cal H}_{n_{j}}(x)$ are the Hermite polynomial with integer $%
n_{j}=0,1,\ldots $. If $H_{j}=0,$ the discrete quantum numbers $n_{j}$ have
to be replaced by the momenta $p_{2j},$ and the corresponding function has
the form 
\[
\phi _{p_{2j-1},p_{2j}}(x_{2j-1},x_{2j})=(2\pi )^{-1}\exp \left\{ -i\left(
p_{2j-1}x^{2j-1}+p_{2j}x^{2j}\right) \right\} . 
\]
The symbol $\ln \left( \mp \tilde{\pi}_{-}\right) $ means the principal
branch of the logarithm, $\ln \left( \mp \tilde{\pi}_{-}\right) =\ln \left| 
\tilde{\pi}_{-}\right| +i\pi \Theta \left( \pm \tilde{\pi}_{-}\right) ,$
while the integration paths in the $\tau $-plane, as well as the arguments $%
\pi _{-},$ are shown in FIG.\ref{f1} and FIG.\ref{f2} for the functions $%
_{+}K\left( x_{-}\right) ,$ $_{+}J\left( x_{-}\right) $ and $^{-}K\left(
x_{-}\right) ,$ $^{-}J\left( x_{-}\right) $ respectively. 
\begin{figure}[h]
\begin{picture}(210,150)
\put(0,70){\vector(1,0){200}}
\put(195,77){Re $\tau$}
\put(100,0){\vector(0,1){140}} 
\put(107,133){Im $\tau$}
{\thicklines
\put(100,70){\oval(50,50)[t]}
\put(107,55){O}
\put(25,70){\vector(-1,0){5}}
\put(25,70){\line(1,0){50}}
\put(125,70){\line(1,0){50}}
}
\put(100,70){\circle*{5}}
\put(180,55){$+\infty$}
\put(10,77){$\infty e^{i\pi}$}
\put(25,55){$\pi_-$}
\end{picture}
\caption[f1]{\label{f1}{Pass of integration in ${}_+K(x_-)$ and 
${}_+J(x_-)$} integrals.}
\end{figure}

\begin{figure}[h]
\begin{picture}(210,150)
\put(0,70){\vector(1,0){200}}
\put(195,77){Re $\tau$}
\put(100,0){\vector(0,1){140}} 
\put(107,133){Im $\tau$}
{\thicklines
\put(100,70){\oval(50,50)[b]}
\put(107,77){O}
\put(175,70){\vector(1,0){5}}
\put(25,70){\line(1,0){50}}
\put(125,70){\line(1,0){50}}
}
\put(100,70){\circle*{5}}
\put(195,55){$+\infty$}
\put(10,77){$\infty e^{-i\pi}$}
\put(175,55){$\pi_-$}
\end{picture}
\caption[f2]{\label{f2}{Pass of integration in ${}^-K(x_-)$ and 
${}^-J(x_-)$} integrals.}
\end{figure}
\noindent One assumes the functions $f_{k}(x_{-})$ obey the requirement that 
$_{+}^{-}K\left( x_{-}\right) $ and $_{+}^{-}J\left( x_{-}\right) $ are
analytic functions, and in particular, the functions $f_{k}(x_{-})$ vanish
at $x_{-}\rightarrow \pm \infty $ quite rapidly.

The solution has a different form if $d$ is odd, ${\cal E}=0$, and all the
imaginary eigenvalues of $F_{\mu \nu }$ are not equal to zero (in this case
a plane-wave field is absent), 
\begin{eqnarray}
\ {}_{+}^{-}\phi _{n,\xi ,r}(x) &=&\;{}_{+}^{-}\phi _{n,r}(x^{0})\phi _{n,r}(%
{\bf x})v_{\xi ,r},  \label{w10a} \\
\ \phi _{n,r}({\bf x}) &=&\phi _{p_{1},n_{1}}(x_{1},x_{2})\phi
_{p_{3},n_{2}}(x_{3},x_{4})...\phi _{p_{d-2},n_{(d-1)/2}}(x_{D-1},x_{D})\;\;,
\nonumber \\
\;{}_{+}^{-}\phi _{n,r}(x^{0}) &=&c\exp \left( \pm i\left| _{\mp
}\varepsilon _{nr}\right| x^{0}\right) ,\smallskip \ \left| _{\mp
}\varepsilon _{nr}\right| =\sqrt{m^{2}+\sum_{j=1}^{(d-1)/2}\omega _{j}}, 
\nonumber
\end{eqnarray}
where $c$ is a normalization constant.

Here $v_{\xi ,r}$ are some constant orthonormal spinors, $v_{\xi
,r}^{\dagger }v_{\xi ,r^{\prime }}=\delta _{r,r^{\prime }}.$ Equation (\ref
{e3}) allows one to subject these spinors to some supplementary conditions, 
\begin{equation}
\ \Xi _{\pm }v_{\mp 1,r}=0,\;\Xi _{\pm }=\frac{1}{2}(1\pm \gamma ^{0}\gamma
^{D}),\;{\rm rank}\;\Xi _{\pm }=J_{(d)}=2^{[\frac{d}{2}]-1};  \label{w12}
\end{equation}
\begin{equation}
R_{j}v_{\xi ,r}=r_{j}v_{\xi ,r},\smallskip \ d\geq 4,\smallskip \
R_{j}=i\gamma ^{2j-1}\gamma ^{2j}.  \label{w13}
\end{equation}
If $d=2,3$, the independent quantum number $r$ does not appear.

One can verify the solutions of the Dirac equation with different $\xi $,
namely, $\left( {\cal P}_{\mu }\gamma ^{\mu }+m\right) \ {}_{+}\phi
_{p_{-},n,+1,r}(x)$ and $\left( {\cal P}_{\mu }\gamma ^{\mu }+m\right) \
{}_{+}\phi _{p_{-},n,-1,r}(x),$ or $\left( {\cal P}_{\mu }\gamma ^{\mu
}+m\right) \ {}^{-}\phi _{p_{-},n,+1,r}(x)$ and $\left( {\cal P}_{\mu
}\gamma ^{\mu }+m\right) \ {}^{-}\phi _{p_{-},n,-1,r}(x)$ are linearly
dependent for each sign ''$+$'' or ''$-$''. It means, in fact, that the spin
projections of a particle (+) and an antiparticle ($-$) can take on only $%
J_{(d)}$ values. Thus, to construct the complete sets it is sufficient to
use only the following solutions: 
\begin{equation}
{}\ {}_{+}^{-}\psi _{p_{-},n,r}(x){}=\left( {\cal P}_{\mu }\gamma ^{\mu
}+m\right) \ {}_{+}^{-}\phi _{p_{-},n,+1,r}(x){}\;.  \label{w14}
\end{equation}
In particular, if $d=2,3$, there is only one spin projection, with only one
spinor $v_{+1,r}=\frac{1}{\sqrt{2}}\left( 
\begin{array}{c}
1 \\ 
i
\end{array}
\right) $. To get the case where $E=0$ one has to consider a limit in the
solution ${}\ {}_{+}^{-}\psi _{p_{-},n,r}(x){}$, such that $-\left(
qE\right) ^{-1}\ln \left( \mp \tilde{\pi}_{-}\right) \rightarrow x_{-}/p_{-}$%
, and $p_{-}>0$ for $\{{}_{+}\psi _{p_{-},n,r}(x)\}$ whereas $p_{-}<0$ for $%
\{{}^{-}\psi _{p_{-},n,r}(x)\}$.

Using the solutions (\ref{w10a}) we may construct the Dirac spinors in a
slightly different way, 
\begin{eqnarray}
_{+}^{-}\psi _{n,\bar r}(x){}\; &=&\left( {\cal P}_\mu \gamma ^\mu +m\right)
_{+}^{-}\phi _{n,\bar r}(x){}\;,  \label{w14b} \\
_{+}^{-}\phi _{n,\bar r}(x){}\; &=&\left( _{+}^{-}\phi _{n,+1,r}(x){}\;+{\rm %
sgn}\left( qH_{(d-1)/2}\right) _{+}^{-}\phi _{n,-1,r}(x){}\;\right)
{},{}\smallskip\ \bar r=\left( r_1,r_{2,}\ldots ,r_{(d-3)/2},{\rm sgn}\left(
qH_{(d-1)/2}\right) \right) ,  \nonumber
\end{eqnarray}
where 
\[
R_j\smallskip\ _{+}^{-}\phi _{n,\bar r}(x){}\;=\bar r_j\smallskip\
_{+}^{-}\phi _{n,\bar r}(x){},\smallskip\ j=1,\;2,\ldots ,(d-1)/2\;, 
\]
and $_{+}^{-}\phi _{n,\bar r}(x){}$ are eigenvectors of the time-independent
operator $m^2-{\cal P}^i{\cal P}_i+ \frac q2\sigma ^{\mu \nu }F_{\mu \nu }.$

The solutions $_{+}\psi _{n,\bar{r}}(x){}$ from (\ref{w14b}) describe
particles (+) and $^{-}\psi _{n,\bar{r}}(x)$ describe antiparticles ($-$) in
any time instant. They form a complete and orthonormal set of solutions.
There is an interesting asymmetry in energy spectrums of the such particles
and antiparticles in odd dimensions. If all quantum numbers $n_{j}=0$, and $%
r_{j}={\rm sgn}\left( qH_{j}\right) ,$ then ${\bf \gamma P}_{+}^{-}\psi _{n,%
\bar{r}}(x){}=0,$ and the ground state of the particle has energy $%
_{+}\varepsilon _{0}=m$, but the energy of the ground state of the
antiparticle is dependent of the magnetic field and different: $\left|
^{-}\varepsilon _{0}\right| =\sqrt{m^{2}+2\min \left| qH_{j}\right| }$.

In the $d=4$ case the solutions similar to (\ref{w14}) were found first in 
\cite{BGJ}. In case $E=0$ they coincide with ones in \cite{R}, and if $H=0$
they coincide with well-known Wolkov form \cite{W}. In the case $E=0,$ the
solutions form a complete and orthonormal set, moreover $_{+}\psi
_{p_{-},n,r}(x){}$ describe particles and $^{-}\psi _{p_{-},n,r}(x){}$
describe antiparticles in any time instant. It is no longer if $E\neq 0$.
However, solutions (\ref{w14}) can be used to construct new complete sets
which can be classified by the correct way.

One can form two complete and orthonormal sets of the solutions:$\left\{
{}_{\pm }\psi _{p_{-},n,r}(x){}\right\} $ and ${}^{\pm }\psi
_{p_{-},n,r}(x){}$ using (\ref{w14}) and additional sets 
\begin{eqnarray}
{}^{+}\psi _{p_{-},n,r}(x) &=&\Theta (\pi _{-})\left( _{+}\psi (x)G\left(
{}_{+}|{}^{+}\right) \right) _{p_{-},n,r}\;,  \nonumber \\
{}_{-}\psi _{p_{-},n,r}(x) &=&\Theta (-\pi _{-})\left( ^{-}\psi (x)G\left(
{}_{-}|{}^{-}\right) ^{\dagger }\right) _{p_{-},n,r}\;,  \label{w14a}
\end{eqnarray}
where $G\left( {}_{\zeta }|{}^{\zeta ^{\prime }}\right) $ obey the relations
(\ref{w15}) and, in particular, $G\left( {}_{+}|{}^{-}\right) $ are
decomposition coefficients of ${}\ {}^{-}\psi _{p_{-},n,r}(x)$ ${}$solutions
in ${}\ {}_{+}\psi _{p_{-},n,r}(x){}$ solutions, 
\begin{equation}
G\left( {}_{+}|{}^{-}\right) _{p_{-},n,r,p_{-}^{\prime },n^{\prime
},r^{\prime }}=({}_{+}\psi _{p_{-},n,r},{}^{-}\psi _{p_{-}^{\prime
},n^{\prime },r^{\prime }}).\;  \label{w16}
\end{equation}
The corresponding Klein-Gordon solutions follow from (\ref{w10}) by the
replacement $u_{\xi ,r}(x_{-})=\exp \left\{ -\frac{1}{2}\ln \left( \mp 
\tilde{\pi}_{-}\right) \right\} .$

The orthonormality of all the solutions can be verified by using the
following integral transformations: 
\begin{eqnarray}
&&{}_{\pm }\psi _{p_{-},n,r}(x)=(2\pi qE)^{-1/2}\int_{-\infty }^{+\infty
}M^{\ast }(p_{D},p_{-}){}_{\pm }\psi _{p_{D},n,r}(x)dp_{D},  \label{w17} \\
&&{}_{\pm }\psi _{p_{D},n,r}(x)=(2\pi qE)^{-1/2}\int_{-\infty }^{+\infty
}M(p_{D},p_{-}){}_{\pm }\psi _{p_{-},n,r}(x)dp_{-},  \label{w18} \\
&&M(p_{D},p_{-})=\exp \left\{ -\frac{i}{4qE}%
((p_{-}-2p_{D})^{2}-2(p_{D})^{2})\right\} ,  \nonumber \\
&&\int_{-\infty }^{+\infty }M^{\ast }(p_{D},p_{-}^{\prime
})M(p_{D},p_{-})dp_{D}=2\pi qE\delta (p_{-}-p_{-}^{\prime }).  \nonumber
\end{eqnarray}
The same relation is valid for functions with ($\pm $) indices above.

The saddle points $\pi _{-}=-2(qEx_{0}-p_{D})$ give the main contribution to
the integrals (\ref{w18}) with ${}_{+}^{-}\psi _{p_{-},n,r}(x)$ functions at 
$x_{0}\rightarrow \pm \infty $ (this was first found in Ref. \cite{NaNi} for
the $d=4$ case without magnetic field). Since the plane wave vanishes in $%
{}_{+}^{-}\psi _{p_{-},n,r}(x)$ at $\pi _{-}\rightarrow \pm \infty ,$
relations (\ref{w18}) reduce at $x_{0}\rightarrow \pm \infty $ to the
space-time uniform field case and can be presented by use of formulas from 
\cite{TIT} as follows 
\begin{eqnarray}
{}\ {}_{+}^{-}\psi _{p_{D},n,r}(x){} &=&\left( {\cal P}_{\mu }\gamma ^{\mu
}+m\right) \ {}_{+}^{-}\phi _{p_{D},n,r}(x){}\;,\smallskip \
x_{0}\rightarrow \pm \infty ,  \label{w19} \\
_{+}^{-}\phi _{p_{D},n,r}(x){} &=&\;{}_{+}^{-}\phi _{n,r}(x^{0})\exp \left(
-ip_{D}x^{D}\right) \ \phi _{n,r}(x_{\perp })v_{+1,r},  \nonumber \\
{}{}_{+}^{-}\phi _{n,r}(x^{0}) &=&CD_{\nu -1}[\pm (1-i)\xi ],\;  \nonumber
\end{eqnarray}

\[
\xi =\left( qEx^{0}-p_{D}\right) /\sqrt{qE},\smallskip \ \nu =i\lambda
/2,\smallskip \ C=(4\pi qE)^{-1/2}\exp \{(-\pi /2+i\ln 2)\lambda /4\}(-i), 
\]
where the function $\ \phi _{n,r}(x_{\perp })$ was defined in (\ref{w10}),
and, $D_{\nu }(z)$ is the Weber parabolic cylinder function \cite{HTF}. Such
solutions were discussed in \cite{GG1996a}. One can now verify the
orthonormality relations of the ${}$sets$\left\{ {}_{+}\psi
_{p_{D},n,r}(x){}\right\} $ and $\left\{ {}^{-}\psi
_{p_{D},n,r}(x){}\right\} .$ Using transformation (\ref{w17}) one gets the
orthonormality relations of sets $\left\{ {}_{+}\psi
_{p_{-},n,r}(x){}\right\} $ and $\left\{ {}^{-}\psi
_{p_{-},n,r}(x){}\right\} $ as well. Using the explicit forms of solutions (%
\ref{w14}), (\ref{w14a}), and relations (\ref{w15}) one can derive the
following relations, 
\begin{eqnarray}
{}^{+}\psi (x)G\left( {}_{+}|{}^{+}\right) ^{\dagger } &=&_{+}\psi
(x)-{}^{-}\psi (x)G\left( {}_{+}|{}^{-}\right) ^{\dagger },  \nonumber \\
{}_{-}\psi (x)G\left( {}_{-}|{}^{-}\right) &=&^{-}\psi (x)-_{+}\psi
(x)G\left( {}_{+}|{}^{-}\right) .  \label{w20}
\end{eqnarray}
By means of the latter one can get the orthonormality relations for both
sets of the solutions, 
\begin{eqnarray}
&&\left( _{\zeta }\psi _{p_{-},n,r},\;_{\zeta ^{\prime }}\psi
_{p_{-}^{\prime },n^{\prime },r^{\prime }}\right) =\delta _{\zeta \zeta
^{\prime }}\delta _{rr^{\prime }}\delta _{nn^{\prime }}\delta
(p_{-}-p_{-}^{\prime }),\;\;  \nonumber \\
&&\left( ^{\zeta }\psi _{p_{-},n,r},\;^{\zeta ^{\prime }}\psi
_{p_{-}^{\prime },n^{\prime },r^{\prime }}\right) =\delta _{\zeta \zeta
^{\prime }}\delta _{rr^{\prime }}\delta _{nn^{\prime }}\delta
(p_{-}-p_{-}^{\prime }),  \label{w21}
\end{eqnarray}
where $\zeta ,\zeta ^{\prime }=\pm $, $\delta _{nn^{\prime }}$ is the
Kronecker symbol for the discrete spectrum and the $\delta $-function for
the continuous one. Here $r=r^{\prime }=+1$ if $d=2,3.$

To solve a problem of the ($\pm $) classification of the solutions one needs
to study their asymptotic behavior at $x_{0}\rightarrow \pm \infty .$ From
the asymptotic forms (\ref{w19}) it follows \cite{GG1996a} that the
asymptotic of the quasienergies of these solution: ${}_{+}\varepsilon
=-qEx^{0},$ is positive and ${}^{-}\varepsilon =-qEx^{0}$ is negative. Thus,
the solutions $_{+}\psi _{p_{D},n,r}(x){}$ describe particles at $%
x_{0}\rightarrow -\infty ,$ and the solutions $^{-}\psi _{p_{D},n,r}(x){}$
describe antiparticles at $x_{0}\rightarrow +\infty .$ Since the solutions $%
_{-}\psi _{p_{-},n,r}(x)$ are orthogonal to $_{+}\psi _{p_{-},n,r}(x),$ they
describe antiparticles at $x_{0}\rightarrow -\infty ,$ and solutions $%
^{+}\psi _{p_{-},n,r}(x){}$ describe particles at $x_{0}\rightarrow +\infty $
since they are orthogonal to $^{-}\psi _{p_{-},n,r}(x).$ One can verify this
by taking into account that the main contribution to integrals (\ref{w18})
at $x_{0}\rightarrow \pm \infty $ for ${}_{-}^{+}\psi _{p,r}(x)$ is given by
point $\pi _{-}=0$. In this limit the contribution of the plane wave does
not depend on $x$ and the results of transformation (\ref{w18}) are
proportional to a superposition of the solutions in a constant uniform field 
\cite{HTF}, 
\begin{eqnarray}
{}\ {}_{-}^{+}\psi _{p_{D},n,r}(x){} &=&\sum_{n,r}a_{n^{\prime }r^{\prime
}}\left( {\cal P}_{\mu }\gamma ^{\mu }+m\right) \ {}_{-}^{+}\phi
_{p_{D},n^{\prime },r^{\prime }}(x){}\;,\smallskip \ x_{0}\rightarrow \pm
\infty ,  \label{w22} \\
_{-}^{+}\phi _{p_{D},n,r}(x){} &=&\;{}_{-}^{+}\phi _{n,r}(x^{0})\exp \left(
-ip_{D}x^{D}\right) \ \phi _{n,r}(x_{\perp })v_{+1r},  \nonumber \\
{}{}_{-}^{+}\phi _{n,r}(x^{0}) &=&CD_{-\nu }[\pm (1+i)\xi ],\;  \nonumber
\end{eqnarray}
where $a_{nr\text{ }}$ are some coefficients dependent on the plane-wave
form. From the asymptotic representations (\ref{w22}) one can see \cite
{GG1996a} that the asymptotic of the quasienergies of these solutions: $%
{}^{+}\varepsilon =qEx^{0}$ is positive and ${}_{-}\varepsilon =qEx^{0}$ is
negative.

Both sets $\{{}_{\pm }\psi _{p_{-},n,r}(x)\}$ and $\{{}^{\pm }\psi
_{p_{-},n,r}(x)\}$ are orthonormal and complete at any time instant. The
form of the commutation function, which will be present in the next section,
can serve as direct proof of the last statement. In $d=4$ such complete sets
of the solutions were first found in \cite{GGS}. If the plane wave is absent
and $E=0$ one can get the solutions with defined energies from ${}\
{}_{+}^{-}\psi _{p_{D},n,r}(x)$ (\ref{w19}) considering the next limit in $%
{}_{+}^{-}\phi _{n,r}(x^{0})$: 
\begin{eqnarray*}
{}_{+}^{-}\phi _{n,r}(x^{0}) &\rightarrow &\left[ _{\mp }\varepsilon
_{nr}(_{\mp }\varepsilon _{nr}+p_{D})\right] ^{-1}\exp \left( \pm i\left|
_{\mp }\varepsilon _{nr}\right| x^{0}\right) , \\
\left| _{\mp }\varepsilon _{nr}\right| &=&\sqrt{m^{2}+\sum_{j=1}^{\left[
(d-1)/2\right] }\omega _{j}+\omega _{0}^{\prime }},\smallskip \ \omega
_{0}^{\prime }=\left\{ 
\begin{array}{c}
p_{D}^{2},\smallskip \ d\smallskip \ {\rm is}\smallskip \ {\rm even} \\ 
0,\smallskip \ d\smallskip \ {\rm is}\smallskip \ {\rm odd}
\end{array}
,\right.
\end{eqnarray*}
where $\omega _{j}$ is defined in (\ref{w11}). In this case one can choose $%
^{+}\psi _{p_{D},n,r}(x)=_{+}\psi _{p_{D},n,r}(x)$ and $_{-}\psi
_{p_{D},n,r}(x)=^{-}\psi _{p_{D},n,r}(x).$

Let us select some important properties of the solution. One can see that
the matrix elements $G\left( {}_{\zeta }|{}^{\zeta ^{\prime }}\right) $ are
diagonal with respect to continuous quantum numbers and spin quantum
numbers, 
\begin{equation}
G\left( {}_{\zeta }|{}^{\zeta ^{\prime }}\right) _{p_{-},n,r,p_{-}^{\prime
},n^{\prime },r^{\prime }}=\delta _{rr^{\prime }}\delta (p_{-}-p_{-}^{\prime
})\delta (p_{1}-p_{1}^{\prime })\ldots \delta (p_{2\left[ (d-1)/2\right]
-1}-p_{2\left[ (d-1)/2\right] -1}^{\prime })\;g\left( {}_{\zeta }|{}^{\zeta
^{\prime }}\right) _{nn^{\prime }}\;.  \label{w22a}
\end{equation}
The solutions $_{+}^{-}\psi _{p_{-},n,r}(x){}$ satisfy the orthonormality
conditions on the null-plane, 
\begin{equation}
\int {}_{+}^{-}\bar{\psi}_{p_{-},n,r}(x)\Xi _{-}{}\smallskip \ _{+}^{-}\psi
_{p_{-}^{\prime },n^{\prime },r^{\prime
}}(x){}dx_{+}dx^{1}...dx^{D-1}=\delta _{rr^{\prime }}\delta _{nn^{\prime
}}\delta (p_{-}-p_{-}^{\prime }),\;\mp \pi _{-}>0.  \label{w22b}
\end{equation}
Since $^{+}\psi _{p_{-},n,r}(x){}=0$ for $\pi _{-}<0$ and relations (\ref
{w20}) take place, one gets the following representations for $G\left(
_{+}|^{-}\right) G\left( {}_{+}|{}^{-}\right) ^{\dagger }$ and $G\left(
_{+}|^{-}\right) ^{\dagger }G\left( {}_{+}|{}^{-}\right) $ matrices:

\begin{eqnarray}
\left( G\left( {}_{+}|{}^{-}\right) G\left( {}_{+}|{}^{-}\right) ^{\dagger
}\right) _{p_{-},n,r,p_{-}^{\prime },n^{\prime },r^{\prime }} &=&\int {}_{+}%
\bar{\psi}_{p_{-},n,r}(x)\Xi _{-}{}\smallskip \ _{+}\psi _{p_{-}^{\prime
},n^{\prime },r^{\prime }}(x){}dx_{+}dx^{1}...dx^{D-1},\;\pi _{-}<0, 
\nonumber \\
\left( G\left( {}_{+}|{}^{-}\right) ^{\dagger }G\left( {}_{+}|{}^{-}\right)
\right) _{p_{-},n,r,p_{-}^{\prime },n^{\prime },r^{\prime }} &=&\int {}^{-}%
\bar{\psi}_{p_{-},n,r}(x)\Xi _{-}{}\smallskip \ ^{-}\psi _{p_{-}^{\prime
},n^{\prime },r^{\prime }}(x){}dx_{+}dx^{1}...dx^{D-1},\;\pi _{-}>0.
\label{w22c}
\end{eqnarray}
Thus one can calculate $_{+}{\cal D}=g\left( _{+}|^{-}\right) g\left(
{}_{+}|{}^{-}\right) ^{\dagger }$ and $^{-}{\cal D}=g\left( _{+}|^{-}\right)
^{\dagger }g\left( {}_{+}|{}^{-}\right) $ matrices using the following
integrals with the solutions of the squared Dirac equation: 
\begin{equation}
_{+}^{-}{\cal D}_{nn^{\prime }}=\int {}\ {}_{+}^{-}\phi
_{p_{-},n,+1,r}^{\dagger }(x)\ {}_{+}^{-}\phi _{p_{-},n^{\prime
},+1,r}(x){}dx^{2}dx^{4}...dx^{2\left[ (d-1)/2\right] },\;\mp \pi _{-}<0.
\label{w22d}
\end{equation}

\section{Green functions}

The perturbation theory with respect to the radiative interaction for the
matrix elements of the processes also has the usual Feynman structure in an
external field creating pairs \cite{GG1977,FG,FGS}. The Feynman diagrams
have to be calculated by means of the causal propagator 
\begin{equation}
\ S^{c}(x,x^{\prime })=c_{v}^{-1}i<0,out|T\psi (x)\bar{\psi}(x^{\prime
})|0,in>,\;\;c_{v}=<0,out|0,in>\;,  \label{w23}
\end{equation}
where $\psi (x)$ is quantum spinor field in the generalized Furry
representation, satisfying the Dirac equation (\ref{e1}), $|0,in>$ and $%
|0,out>$ are the initial and the final vacuum in this representation, and $%
c_{v}$ is the vacuum to vacuum transition amplitude. The propagator $\
S^{c}(x,x^{\prime })$ obeys the equation 
\begin{equation}
\left( {\cal P}_{\mu }\gamma ^{\mu }-m\right) \ S^{c}(x,x^{\prime })=-\delta
^{(d)}(x-x^{\prime })\;,  \label{w24}
\end{equation}
and is a Green function of the equation. Another important singular function
is the commutation function 
\begin{equation}
S(x,x^{\prime })=i\left[ \psi (x),\bar{\psi}(x^{\prime })\right] _{+}.
\label{w23a}
\end{equation}
It obeys the homogeneous Dirac equation (\ref{e1}) and the initial condition 
\begin{equation}
\left. S(x,x^{\prime })\right| _{x_{0}=x_{0}^{\prime }}=i\gamma ^{0}\delta (%
{\bf x}-{\bf x}^{\prime }).  \label{w23b}
\end{equation}
The commutation function $S(x,x^{\prime })$ is at the same time the
propagation function of the Dirac equation, i.e. it connects solutions of
the equation in two different time instants.

QED with unstable vacuum has a number of peculiarities. Thus, for instance,
in the calculation of the expectation values and the total probabilities
Green functions of different types from (\ref{w23}) appear \cite
{GG1977,FG,FGS}: 
\begin{eqnarray}
S_{in}^{c}(x,x^{\prime }) &=&i<0,in|T\psi (x)\bar{\psi}(x^{\prime })|0,in>, 
\nonumber \\
S_{in}^{-}(x,x^{\prime }) &=&i<0,in|\psi (x)\bar{\psi}(x^{\prime })|0,in>, 
\nonumber \\
S_{in}^{+}(x,x^{\prime }) &=&i<0,in|\bar{\psi}(x^{\prime })\psi (x)|0,in>, 
\nonumber \\
S_{in}^{\bar{c}}(x,x^{\prime }) &=&i<0,in|\psi (x)\bar{\psi}(x^{\prime
})T|0,in>,  \nonumber \\
S_{out}^{c}(x,x^{\prime }) &=&i<0,out|T\psi (x)\bar{\psi}(x^{\prime
})|0,out>,  \label{w25}
\end{eqnarray}
where the symbol of the $T$-product acts on both sides: it orders the field
operators to the right of its and antiorders them to the left. The functions 
$S_{in}^{c}(x,x^{\prime }),$ $S_{out}^{c}(x,x^{\prime })$ obey Eq. (\ref{w24}%
), $S^{\mp }(x,x^{\prime })$ satisfies Eq. (\ref{e1}) and $S_{in}^{\bar{c}%
}(x,x^{\prime })$ obeys the equation 
\begin{equation}
\left( {\cal P}_{\mu }\gamma ^{\mu }-m\right) \ S_{in}^{\bar{c}}(x,x^{\prime
})=\delta ^{(d)}(x-x^{\prime })\;.  \label{w26}
\end{equation}
As well, all these different kinds of the Green functions are used to
represent various matrix elements of operators of the current and
energy-momentum tensor, and effective action beginning with zeroth order
with respect to radiative interaction.

Solutions (\ref{w14}) and (\ref{w14a}) with quantum number $p_{-}$ are
especially adapted to calculate all the Green functions. One can express the
Green functions via solutions (\ref{w14}) and (\ref{w14a}) \cite
{GG1977,FG,FGS}: 
\begin{eqnarray}
S^{c}\left( x,x^{\prime }\right) &=&\theta \left( x_{0}-x_{0}^{\prime
}\right) S^{-}\left( x,x^{\prime }\right) -\theta \left( x_{0}^{\prime
}-x_{0}\right) S^{+}\left( x,x^{\prime }\right) ,  \label{w27} \\
S\left( x,x^{\prime }\right) &=&S^{-}\left( x,x^{\prime }\right)
+S^{+}\left( x,x^{\prime }\right) ,  \label{w28}
\end{eqnarray}
\begin{eqnarray}
S^{-}\left( x,x^{\prime }\right) &=&i\int_{-\infty }^{+\infty
}dp_{-}\sum_{nr\{n_{j}^{\prime }\}}{}^{+}\psi _{p_{-,}n,r}\left( x\right)
g\left( \left. _{+}\right| ^{+}\right) _{nn^{\prime }}^{-1}\smallskip \ _{+}%
\bar{\psi}_{p_{-},n^{\prime },r}\left( x^{\prime }\right) ,  \nonumber
\label{H22} \\
S^{+}\left( x,x^{\prime }\right) &=&i\int_{-\infty }^{+\infty
}dp_{-}\sum_{nr\{n_{j}^{\prime }\}}\smallskip \ _{-}\psi _{p_{-},n,r}\left(
x\right) \left[ g\left( \left. _{-}\right| ^{\_}\right) ^{-1}\right]
_{nn^{\prime }}^{\ast }\smallskip \ ^{-}\bar{\psi}_{p_{-},n^{\prime
},r}\left( x^{\prime }\right) ,  \label{w29}
\end{eqnarray}
\begin{eqnarray}
S_{in}^{c}\left( x,x^{\prime }\right) &=&\theta \left( x_{0}-x_{0}^{\prime
}\right) S_{in}^{-}\left( x,x^{\prime }\right) -\theta \left( x_{0}^{\prime
}-x_{0}\right) S_{in}^{+}\left( x,x^{\prime }\right) ,  \label{w30} \\
S_{in}^{\bar{c}}\left( x,x^{\prime }\right) &=&\theta \left( x_{0}^{\prime
}-x_{0}\right) S_{in}^{-}\left( x,x^{\prime }\right) -\theta \left(
x_{0}-x_{0}^{\prime }\right) S_{in}^{+}\left( x,x^{\prime }\right) ,
\label{w31}
\end{eqnarray}
\begin{equation}
S_{in}^{\mp }\left( x,x^{\prime }\right) =i\int_{-\infty }^{+\infty
}dp_{-}\sum_{nr}\smallskip \ _{\pm }\psi _{p_{-},n,r}\left( x\right) _{\pm }%
\bar{\psi}_{p_{-},n,r}\left( x^{\prime }\right) ,  \label{w32}
\end{equation}
\begin{eqnarray}
S_{out}^{c}\left( x,x^{\prime }\right) &=&\theta \left( x_{0}-x_{0}^{\prime
}\right) S_{out}^{-}\left( x,x^{\prime }\right) -\theta \left( x_{0}^{\prime
}-x_{0}\right) S_{out}^{+}\left( x,x^{\prime }\right) ,  \label{w33} \\
S_{out}^{\mp }\left( x,x^{\prime }\right) &=&i\int_{-\infty }^{+\infty
}dp_{-}\sum_{nr}\smallskip \ ^{\pm }\psi _{p_{-},n,r}\left( x\right) ^{\pm }%
\bar{\psi}_{p_{-},n,r}\left( x^{\prime }\right) .  \label{w34}
\end{eqnarray}
where all $p_{j}^{\prime }=p_{j\text{, }}$the symbol $\sum_{nr}$ means the
summation over all discrete quantum numbers $n_{j},r_{j}$ and the
integration over all continuous $p_{j}$, and the symbol $\sum_{\{n_{j}\}}$
means the summation over all discrete quantum numbers $n_{j}$ only. Using
the relations between the Green functions and between the matrices $G\left(
{}_{\zeta }|{}^{\zeta ^{\prime }}\right) $ one can present the functions $%
S^{\mp },$ $S_{in}^{\mp }$ and $S_{out}^{\mp }$ as follows 
\begin{eqnarray}
\ &&\pm S^{\mp }(x,x^{\prime })=S^{c}(x,x^{\prime })\pm \theta (\mp
(x_{0}-x_{0}^{\prime }))S(x,x^{\prime })\quad ,  \label{w35} \\
\ &&\pm S_{in}^{\mp }(x,x^{\prime })=S_{in}^{c}(x,x^{\prime })\pm \theta
(\mp (x_{0}-x_{0}^{\prime }))S(x,x^{\prime })\quad ,  \label{w36} \\
\ &&\pm S_{out}^{\mp }(x,x^{\prime })=S_{out}^{c}(x,x^{\prime })\pm \theta
(\mp (x_{0}-x_{0}^{\prime }))S(x,x^{\prime })\quad ,  \label{w37} \\
\ &&S_{in}^{c}(x,x^{\prime })=S^{c}(x,x^{\prime })-S^{a}(x,x^{\prime })\quad
,  \label{w38} \\
\ &&S_{out}^{c}(x,x^{\prime })=S^{c}(x,x^{\prime })-S^{p}(x,x^{\prime
})\quad ,  \label{w39} \\
\ &&S^{a}(x,x^{\prime })=-i\int_{-\infty }^{+\infty
}dp_{-}\sum_{nr\{n_{j}^{\prime }\}}\smallskip \ _{-}{\psi }_{p_{-},n,r}(x)\,%
\left[ g(_{+}|^{-})g(_{-}|^{-})^{-1}\right] _{nn^{\prime }}^{\dagger }{_{+}%
\bar{\psi}}_{p_{-},n^{\prime },r}(x^{\prime })\quad ,  \label{w40} \\
\ &&S^{p}(x,x^{\prime })=i\int_{-\infty }^{+\infty
}dp_{-}\sum_{nr\{n_{j}^{\prime }\}}\smallskip \ ^{+}{\psi }_{p_{-},n,r}(x)\,%
\left[ g(_{+}|^{+})^{-1}g(_{+}|^{-})\right] _{nn^{\prime }}{^{-}\bar{\psi}}%
_{p_{-},n^{\prime },r}(x^{\prime })\quad .  \label{w41}
\end{eqnarray}
To calculate all the types of Green functions it is sufficient to take sums
in $S^{\pm }(x,x^{\prime })$ and $S^{a,p}(x,x^{\prime })$ only. That will be
done below.

It follows from (\ref{w14a}) 
\begin{eqnarray}
S^{\pm }(x,x^{\prime }) &=&\int_{-\infty }^{+\infty }\Theta \left( \mp \pi
_{-}\right) _{+}^{-}Y(x,x^{\prime },p_{-})dp_{-},  \label{w42} \\
_{+}^{-}Y(x,x^{\prime },p_{-}) &=&i\sum_{nr}{}_{+}^{-}\psi
_{p_{-,}n,r}\left( x\right) \smallskip \ _{+}^{-}\overline{\psi }%
_{p_{-},n,r}\left( x^{\prime }\right) .  \label{w42a}
\end{eqnarray}
Also, taking into account Eqs. (\ref{w20}) one gets 
\begin{eqnarray}
S^{a}(x,x^{\prime }) &=&-\int_{-\infty }^{+\infty }\Theta \left( -\pi
_{-}\right) _{+}Y(x,x^{\prime },p_{-})dp_{-},  \nonumber \\
S^{p}(x,x^{\prime }) &=&\int_{-\infty }^{+\infty }\Theta \left( \pi
_{-}\right) ^{-}Y(x,x^{\prime },p_{-})dp_{-},  \label{w43}
\end{eqnarray}
Due to the fact that $_{+}\psi _{p_{-,}n,r}\left( x\right) $ and $^{-}\psi
_{p_{-,}n,r}\left( x\right) $ have similar form (\ref{w14}), the sums in (%
\ref{w42a}) can be taken in the same way. Since 
\[
\sum_{r}v_{+1,r}v_{+1,r}^{+}=\Xi _{+}\smallskip \ {\rm if}\smallskip \
d>3,\smallskip \ {\rm and}\smallskip \ v_{+1}v_{+1}^{+}=\Xi _{+}\smallskip \ 
{\rm if}\smallskip \ d=2,3, 
\]
the summation over the spin quantum numbers can be done in (\ref{w42a}) to
get 
\begin{eqnarray}
_{+}^{-}Y(x,x^{\prime },p_{-}) &=&\left[ \gamma ^{0}+\left( \gamma _{\bot }%
{\cal P}_{\bot }+m\right) \frac{1}{\pi _{-}}\right] \Xi _{+}\exp \left( -i%
\frac{q}{2}\sigma ^{\mu \nu }F_{\mu \nu }^{\perp }\smallskip \
_{+}^{-}a\right) \left[ \gamma ^{0}+\left( -\gamma _{\bot }{\cal P^{\prime }}%
_{\bot }^{\ast }+m\right) \frac{1}{\pi _{-}^{\prime }}\right]  \nonumber \\
&&\ \ \ \ \gamma ^{0}\exp \left\{ \frac{1}{2}\left[ \ln \left( \mp \tilde{\pi%
}_{-}\right) +\left( \ln \left( \mp \tilde{\pi}_{-}^{\prime }\right) \right)
^{\ast }\right] \right\} \cdot \sqrt{qE}_{+}^{-}\tilde{f}^{(0)}\left(
x,x^{\prime },p_{-}\right) ,  \label{w44} \\
_{+}^{-}\tilde{f}^{(0)}\left( x,x^{\prime },p_{-}\right)
&=&i\sum_{n}{}_{+}^{-}\varphi _{p_{-,}n,r}\left( x\right) \smallskip \
_{+}^{-}\varphi _{p_{-},n,r}^{\ast }\left( x^{\prime }\right) ,  \label{w45}
\\
_{+}^{-}\varphi _{p_{-,}n,r}\left( x\right) &=&\exp \left\{ -\frac{1}{2}\ln
\left( \mp \tilde{\pi}_{-}\right) \right\} \;{}_{+}^{-}\phi
_{p_{-},n,r}(x^{0},x^{D})\phi _{n,r}(x_{\perp }),  \nonumber \\
\ \ _{+}^{-}a &=&\left( 2qE\right) ^{-1}\left[ \left( \ln \left( \mp \tilde{%
\pi}_{-}^{\prime }\right) \right) ^{\ast }-\ln \left( \mp \tilde{\pi}%
_{-}\right) \right] ,  \label{w46} \\
\smallskip \ \tilde{\pi}_{-}^{\prime } &=&\tilde{\pi}_{-}+\sqrt{qE}%
y_{-},\smallskip \ y_{\mu }=x_{\mu }-x_{\mu }^{\prime },  \nonumber \\
{\cal P}_{\perp \mu } &=&0\smallskip \ {\rm if}\smallskip \ \mu
=0,D,\smallskip \ \ {\cal P}_{\perp \mu }={\cal P}_{\mu }\smallskip \ \ {\rm %
if}\smallskip \ k=1,\ldots ,D-1\;,  \nonumber \\
&&{\cal P^{\prime }}_{\mu }^{\ast }=-i\frac{\partial }{\partial x^{\prime
}{}^{\mu }}-q\,A_{\mu }(x^{\prime }).  \nonumber
\end{eqnarray}
It is convenient to make the replacement 
\[
u=\left( 2qE\right) ^{-1}\left[ \left( \ln \left( \mp \tilde{\pi}%
_{-}^{\prime }\right) \right) ^{\ast }-\ln \left( \mp \tau \right) \right] 
\]
in the functions $_{+}^{-}K(x_{-})$ and $_{+}^{-}J(x_{-})$, and the one 
\[
u=\left( 2qE\right) ^{-1}\left[ \left( \ln \left( \tilde{\pi}_{-}^{\prime
}\right) \right) ^{\ast }-\ln \tau \right] 
\]
in the functions $_{+}^{-}K^{\ast }(x_{-}^{\prime })$ and $_{+}^{-}J^{\ast
}(x_{-}^{\prime })$ (from $_{+}^{-}\phi _{p_{-},n,r}(x^{0},x^{D})$ and $%
_{+}^{-}\phi _{p_{-},n,r}^{\ast }(x^{\prime 0},x^{\prime D})$ defined in (%
\ref{w10})). We do not discuss the integration paths over the variable $u,$
since it is natural to consider the plane-wave potential $f(x_{-})$ to be an
entire function on $x_{-\text{ }}$, in that case the integrals $%
_{+}^{-}K(x_{-})$, $_{+}^{-}J(x_{-})$, $_{+}^{-}K^{\ast }(x_{-}^{\prime })$
and $_{+}^{-}J^{\ast }(x_{-}^{\prime })$ are well defined by the integration
limits. For different forms of the plane-wave potential the integration
paths over $u$ are no longer arbitrary, however, they easily are extracted
from the integration paths shown in Fig.1 and Fig.2. Then, one can present
some combinations involved in (\ref{w45}) in the following way: 
\begin{eqnarray}
_{+}^{-}K\left( x_{-}\right) -\;_{+}^{-}K^{\ast }\left( x_{-}^{\prime
}\right) &=&l\left( _{+}^{-}a\right) +2\left( e^{2qF_{+}^{-}a}-1\right)
\int_{_{+}^{-}b}^{0}e^{-2qFu}qf\left( x_{-}\left( u\right) \right) du, 
\nonumber \\
_{+}^{-}K\left( x_{-}\right) +\;_{+}^{-}K^{\ast }\left( x_{-}^{\prime
}\right) &=&l\left( _{+}^{-}a\right) +2\left( e^{2qF_{+}^{-}a}+1\right)
\int_{_{+}^{-}b}^{0}e^{-2qFu}qf\left( x_{-}\left( u\right) \right) du,
\label{w47} \\
_{+}^{-}J\left( x_{-}\right) -\;_{+}^{-}J^{\ast }\left( x_{-}^{\prime
}\right) &=&\Phi \left( _{+}^{-}a\right) +2\int_{0}^{_{+}^{-}a}qf\left(
x_{-}\left( u\right) \right) e^{2qFu}du\smallskip \
qF\int_{_{+}^{-}b}^{0}e^{-2qFu^{\prime }}qf\left( x_{-}\left( u^{\prime
}\right) \right) du^{\prime },  \nonumber
\end{eqnarray}
where 
\begin{eqnarray}
\Phi \left( _{+}^{-}a\right) &=&\int_{0}^{_{+}^{-}a}qf\left( x_{-}\left(
u\right) \right) \left[ qf\left( x_{-}\left( u\right) \right) +qFl\left(
u\right) \right] du,  \nonumber \\
l\left( u\right) &=&2\int_{0}^{u}e^{2qF\left( u-u^{\prime }\right) }qf\left(
x_{-}\left( u^{\prime }\right) \right) du^{\prime },  \label{w48} \\
_{+}^{-}b &=&-\infty -i\pi /\left( 2qE\right) \Theta \left( \pm \pi
_{-}^{\prime }\right) ,\smallskip \ x_{-}(u)=x_{-}^{\prime }+y_{-}\frac{%
1-\exp \left\{ -2qEu\right\} }{1-\exp \left\{ -2qE_{+}^{-}a\right\} }. 
\nonumber
\end{eqnarray}
The summation over $n_{j}$ in (\ref{w45}) can be performed using the
M\"{o}ller formula \cite{HTF}. After integration over $p_{j}$ in (\ref{w45}%
), applying the operators $\exp \left\{ -i_{+}^{-}K\left( x_{-}\right) \pi
_{\perp }\right\} $ and $\exp \left\{ -i_{+}^{-}K\left( x_{-}\right) \pi
_{\perp }^{\prime }\right\} ^{\ast }$, and using relations (\ref{w47}), we
obtain 
\begin{eqnarray}
\ &&_{+}^{-}\tilde{f}^{(0)}\left( x,x^{\prime },p_{-}\right) =\exp \left\{
iq\Lambda -im^{2}\smallskip \ _{+}^{-}a\right. -  \nonumber \\
&&\ \ \left. -\frac{1}{2}\left[ \ln \left( \mp \tilde{\pi}_{-}\right)
+\left( \ln \left( \mp \tilde{\pi}_{-}^{\prime }\right) \right) ^{\ast }%
\right] -i\frac{\pi _{-}+\pi _{-}^{\prime }}{4}y_{+}\right\} h(_{+}^{-}a\ ),
\label{w49} \\
\ h(_{+}^{-}a\ ) &=&\exp \left\{ i\Phi (_{+}^{-}a)-i\frac{1}{4}%
(y+l(_{+}^{-}a))qF^{\perp }\coth (qF^{\perp }\smallskip \
_{+}^{-}a)(y+l(_{+}^{-}a))+\frac{i}{2}yqF^{\perp }l(s)\right\} Z_{(d)},
\label{w50}
\end{eqnarray}
where 
\begin{eqnarray}
\ Z_{(d)} &=&c_{d}{\prod_{j=1}^{(d-2)/2}\limits}\,\left( \frac{qH_{j}}{\sin
(qH_{j}s)}\right) ,\quad \;\;\;\;\;\;\;d\mbox{ is 
even},  \nonumber \\
\ Z_{(d)} &=&c_{d}s^{-1/2}{\prod_{j=1}^{(d-3)/2}\limits}\,\left( \frac{qH_{j}%
}{\sin (qH_{j}s)}\right) ,\quad d\mbox{ is odd},  \label{w51} \\
\ c_{d} &=&(4\pi )^{-d/2}\,\exp \left\{ -i\pi (d-4)/4\right\} \;,  \nonumber
\end{eqnarray}
and 
\begin{equation}
\ \Lambda =-\int_{x^{\prime }}^{x}\left( A_{\mu }^{E}+A_{\mu }^{H}\right)
dx^{\mu }.  \label{w52}
\end{equation}
Here $A_{\mu }^{E}+A_{\mu }^{H}$ is a potential of the constant uniform
field $F_{\mu \nu },$ and the integral is taken along the line.

Let us remark that after convenient gauge transformation of electric
constant field potentials the functions $_{+}^{-}\tilde{f}^{(0)}\left(
x,x^{\prime },p_{-}\right) $ obeys the Klein-Gordon equation, 
\begin{equation}
\left( \pi _{-}2i\frac{\partial }{\partial x_{-}}-iqE+{\cal P}_{\bot
}^{2}-m^{2}\right) \exp \left\{ -\frac{iqE}{2}\left( \frac{x_{-}^{2}}{2}%
-x_{D}^{2}\right) \right\} \;{}_{+}^{-}\tilde{f}^{(0)}\left( x,x^{\prime
},p_{-}\right) =0.  \label{w53}
\end{equation}
Taking into account the relations 
\begin{eqnarray}
\pi _{\bot }e^{iq\Lambda } &=&e^{iq\Lambda }\left( i\frac{\partial }{%
\partial x_{\perp }}+\frac{1}{2}qFy_{\bot }\right) ,  \nonumber \\
\pi _{\bot }^{\prime \ast }e^{iq\Lambda } &=&e^{iq\Lambda }\left( -i\frac{%
\partial }{\partial x_{\perp }^{\prime }}-\frac{1}{2}qFy_{\bot }\right) ,
\label{w54}
\end{eqnarray}
\begin{eqnarray}
\exp \left( -iq\sigma ^{0D}F_{0D}\smallskip \ _{+}^{-}a\right) &=&\cosh
\left( qE\smallskip \ _{+}^{-}a\right) -i\sigma ^{0D}\sinh \left(
qE\smallskip \ _{+}^{-}a\right) ,  \nonumber \\
\exp \left( -i\frac{q}{2}\sigma ^{\mu \nu }F_{\mu \nu }^{(j)}\smallskip \
_{+}^{-}a\right) &=&\cos \left( qH_{j}\smallskip \ _{+}^{-}a\right)
+iR_{j}\sin \left( qH_{j}\smallskip \ _{+}^{-}a\right) ,  \label{w55}
\end{eqnarray}
where 
\[
F_{\mu \nu }^{(j)}=H_{j}(\delta _{\mu }^{2j}\delta _{\nu }^{2j-1}-\delta
_{\nu }^{2j}\delta _{\mu }^{2j-1}), 
\]
and matrix $R_{j}$ is defined by (\ref{w13}), one can get relation 
\begin{eqnarray}
&&\exp \left( -i\frac{q}{2}\sigma ^{\mu \nu }F_{\mu \nu }^{\perp }\smallskip
\ _{+}^{-}a\right) \gamma \pi _{\bot }^{\prime \ast }\smallskip \ _{+}^{-}%
\tilde{f}^{(0)}\left( x,x^{\prime },p_{-}\right) =  \nonumber  \label{4p20}
\\
&&\left( \gamma \pi _{\bot }+\gamma qF^{\perp }l\left( _{+}^{-}a\right)
\right) \exp \left( -i\frac{q}{2}\sigma ^{\mu \nu }F_{\mu \nu }^{\perp
}\smallskip \ _{+}^{-}a\right) \smallskip \ _{+}^{-}\tilde{f}^{(0)}\left(
x,x^{\prime },p_{-}\right) .  \label{w56}
\end{eqnarray}
By using formulas (\ref{w55}), (\ref{w56}) and Eq. (\ref{w53}) one can
transform expression (\ref{w44}) to the following form, 
\begin{eqnarray}
_{+}^{-}Y(x,x^{\prime },p_{-}) &=&\mp \left( \gamma {\cal P}+m\right)
\smallskip \ _{+}^{-}\tilde{f}\left( x,x^{\prime },p_{-}\right) ,
\label{w57} \\
\ _{+}^{-}\tilde{f}\left( x,x^{\prime },p_{-}\right) &=&\left[ \exp \left( -i%
\frac{q}{2}\sigma ^{\mu \nu }F_{\mu \nu }\smallskip \ _{+}^{-}a\right) \mp 
\frac{1}{2}(\gamma ^{0}-\gamma ^{D})\gamma \int_{0}^{\
_{+}^{-}a}e^{qF(\smallskip \ _{+}^{-}a-2u)}q\frac{df(x_{-}(u))}{du}du\right.
\nonumber \\
&&\left. \exp \left\{ -\frac{1}{2}\left[ \ln \left( \mp \tilde{\pi}%
_{-}\right) +\left( \ln \left( \mp \tilde{\pi}_{-}^{\prime }\right) \right)
^{\ast }\right] \right\} \right] \smallskip \ _{+}^{-}\tilde{f}^{(0)}\left(
x,x^{\prime },p_{-}\right) .  \label{w58}
\end{eqnarray}

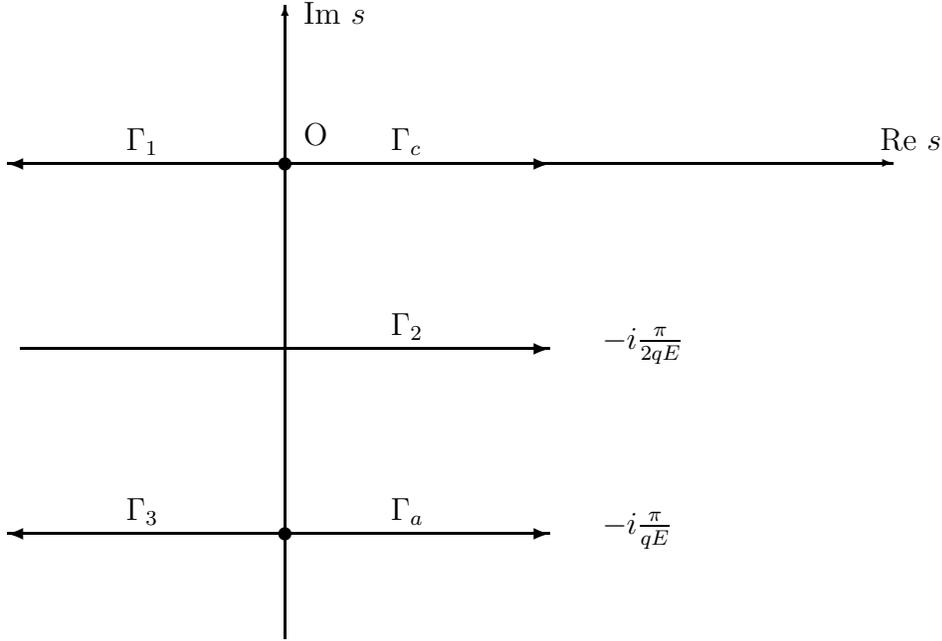
\begin{figure}[h]
\begin{picture}(350,245)
\put(0,180){\vector(1,0){330}}
\put(325,185){Re $s$}
\put(100,0){\vector(0,1){240}} 
\put(107,233){Im $s$}
{\thicklines
\put(0,180){\vector(1,0){200}}
\put(0,110){\vector(1,0){200}}
\put(0,40){\vector(1,0){200}}
\put(0,180){\vector(-1,0){5}}
\put(0,40){\vector(-1,0){5}}
}
\put(140,185){$\Gamma_c$}
\put(140,115){$\Gamma_2$}
\put(140,45){$\Gamma_a$}
\put(40,185){$\Gamma_1$}
\put(40,45){$\Gamma_3$}
\put(220,110){$-i\frac{\pi}{2qE}$}
\put(220,40){$-i\frac{\pi}{qE}$}
\put(100,180){\circle*{5}}
\put(100,40){\circle*{5}}
\put(107,187){O}
\end{picture}
\caption[f3]{\label{f3}{Contours of integration 
$\Gamma_1,\Gamma_2,\Gamma_3,\Gamma_c,\Gamma_a$}.}
\end{figure}

\begin{figure}[h]
\begin{picture}(210,245)
\put(0,170){\vector(1,0){200}}
\put(195,177){Re $s$}
\put(100,0){\vector(0,1){240}} 
\put(107,233){Im $s$}
{\thicklines
\put(100,170){\oval(50,50)[b]}
\put(122,138){$\Gamma$}
\put(85,170){\vector(1,0){5}}
\put(75,170){\line(1,0){50}}
}
%
\put(100,170){\circle*{5}}
\put(107,177){O}
{\thicklines
\put(100,40){\oval(50,50)[t]}
\put(122,60){$\Gamma^a_1$}
\put(115,40){\vector(-1,0){5}}
\put(75,40){\line(1,0){50}}
\put(100,40){\circle*{5}}
\put(150,40){$-i\frac{\pi}{qE}$}}
\end{picture}
\caption[f4]{\label{f4}{Contours of integration $\Gamma, \Gamma^a_1$}.}
\end{figure}
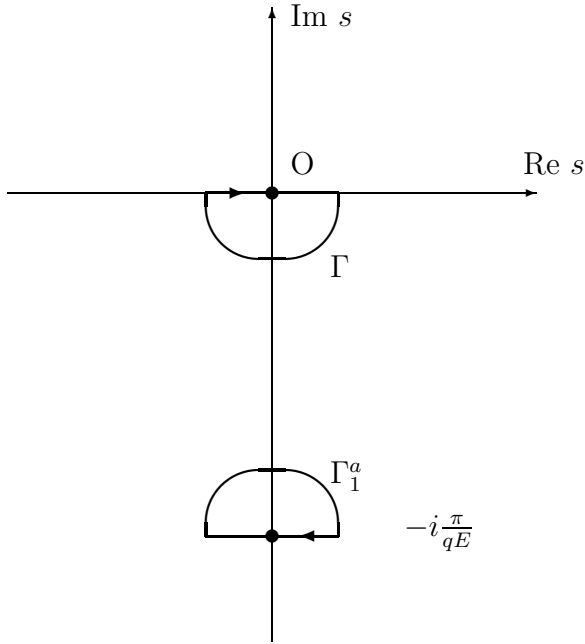

In the external field under consideration the real proper time $S$ is a
function of $\pi _{-}$ and $\pi _{-}^{\prime }$ because the classical
equation of motion has a form $\pi _{-}^{-1}dx_{-}=m^{-1}dS$. Thus if $%
y_{-}\neq 0$ one can transform the $p_{-}$ integration in Green functions
into integration over the Fock-Schwinger proper time by making the change of
the variable, 
\begin{equation}
s={}_{+}^{-}a\;.  \label{w59}
\end{equation}
Then, one gets the following representations for the Green functions: 
\begin{eqnarray}
S^{\mp ,a,p}(x,x^{\prime }) &=&(\gamma {\cal P}+m)\Delta ^{\mp
,a,p}(x,x^{\prime })\;,\;  \nonumber \\
\mp \Delta ^{\pm }(x,x^{\prime }) &=&\int_{\Gamma _{c}}f(x,x^{\prime
},s)ds-\Theta (\pm y_{-})\int_{\Gamma _{c}-\Gamma _{2}-\Gamma
_{1}}f(x,x^{\prime },s)ds\;,  \label{w60} \\
\Delta ^{a}(x,x^{\prime }) &=&\int_{\Gamma _{a}}f(x,x^{\prime },s)ds+\Theta
(y_{-})\int_{\Gamma _{2}+\Gamma _{3}-\Gamma _{a}}f(x,x^{\prime },s)ds\;.
\label{w61} \\
\Delta ^{p}(x,x^{\prime }) &=&\int_{\Gamma _{a}}f(x,x^{\prime },s)ds+\Theta
(-y_{-})\int_{\Gamma _{1}^{a}}f(x,x^{\prime },s)ds\;.  \label{w62}
\end{eqnarray}
All the contours of the integrals are shown on Fig. \ref{f3}. The contours $%
\Gamma _{c}$ and $\Gamma _{1}$ are placed below the singular points on the
real axis everywhere outside of the origin, and 
\begin{eqnarray}
\  &&f(x,x^{\prime },s)=\left[ \exp \left( -i\frac{q}{2}\sigma ^{\mu \nu
}F_{\mu \nu }s\right) \right.   \nonumber \\
&&\left. +(n\gamma )\gamma \int_{0}^{s}e^{qF(s-2u)}\frac{df((nx(u)))}{du}du%
\frac{\sinh (qEs)}{E(ny)}\right] f^{(0)}(x,x^{\prime },s),  \label{w63} \\
\  &&f^{(0)}(x,x^{\prime },s)=\exp \{iq\Lambda \}\frac{qE}{\sinh (qEs)}\exp
\{-im^{2}s+i\Phi (s)  \nonumber \\
&&\ -i\frac{1}{4}(y+l(s))qF\coth (qFs)(y+l(s))+\frac{i}{2}yqFl(s)\}Z_{(d)},
\label{w64}
\end{eqnarray}
where $Z_{(d)}$ is defined in (\ref{w51}). The function $%
f^{(0)}(x,x^{\prime },s)$ has two singular points on the
\noindent  complex region
between the contours $\Gamma _{c}-\Gamma _{1}$ and $\Gamma _{a}-\Gamma _{3}.$
They are situated at the imaginary axis: $s_{0}=0,$ and $qEs_{1}=-i\pi $.
One can transform the contours $\Gamma _{c}-\Gamma _{2}-\Gamma _{1}$ into $%
\Gamma $ and $\Gamma _{2}+\Gamma _{3}-\Gamma _{a}$ into $\Gamma _{1}^{a}$
(see Fig. \ref{f4}) with radii tending to zero. Since 

\[
\int_{\Gamma }f(x,x^{\prime },s)ds=0\;\mbox{if}\;y_{\mu }y^{\mu
}<0\;,\;\;\int_{\Gamma _{1}^{a}}f(x,x^{\prime },s)ds=0\;\mbox{if}%
\;y_{0}^{2}>y_{D}^{2}\;,
\]
one can rewrite (\ref{w60}),(\ref{w61}) and (\ref{w62}) as follows 
\begin{eqnarray}
S^{\mp }(x,x^{\prime }) &=&(\gamma {\cal P}+m)\Delta ^{\mp }(x,x^{\prime
}),\;  \label{w65} \\
\mp \Delta ^{\pm }(x,x^{\prime }) &=&\int_{\Gamma ^{c}}f(x,x^{\prime
},s)ds-\Theta (\pm y_{0})\int_{\Gamma }f(x,x^{\prime },s)ds\;,  \nonumber \\
S^{a,p}(x,x^{\prime }) &=&(\gamma {\cal P}+m)\Delta ^{a,p}(x,x^{\prime }),
\label{w66} \\
\Delta ^{a}(x,x^{\prime }) &=&\int_{\Gamma ^{a}}f(x,x^{\prime },s)ds+\Theta
(-y^{D})\int_{\Gamma _{1}^{a}}f(x,x^{\prime },s)ds\;,  \nonumber \\
\Delta ^{p}(x,x^{\prime }) &=&\int_{\Gamma ^{a}}f(x,x^{\prime },s)ds+\Theta
(y^{D})\int_{\Gamma _{1}^{a}}f(x,x^{\prime },s)ds\;.  \nonumber
\end{eqnarray}
One can check that these expressions are valid for arbitrary $x$ and $%
x^{\prime }$. To prove this one need to verify that the functions $S^{\pm
}(x,x^{\prime })$ and $S^{a,p}(x,x^{\prime }),$ which are presented via
integrals (\ref{w65}) and (\ref{w66}), and by means of representations (\ref
{w42}) and (\ref{w43}), are the same solutions of the Dirac equation for any 
$x$ and $x^{\prime }$. Thus it is enough, to check first that expressions (%
\ref{w65}) and (\ref{w66}) obey the Dirac equation for any $x$ and $%
x^{\prime }$. Then, one has to prove that the Cauchy conditions for
distributions (\ref{w65}) and (\ref{w66}) coincide at $x_{0}=x_{0}^{\prime }$
with (\ref{w42}) and (\ref{w43}) respectively. As to $S^{\pm }(x,x^{\prime
}),$ one can use the proper time representations for $S^{c}(x,x^{\prime })$
and $S(x,x^{\prime })$ functions, which follow from (\ref{w65}), (\ref{w27})
and (\ref{w28}), 
\begin{eqnarray}
S^{c}(x,x^{\prime }) &=&(\gamma {\cal P}+m)\Delta ^{c}(x,x^{\prime }), 
\nonumber \\
\Delta ^{c}(x,x^{\prime }) &=&\int_{\Gamma _{c}}f(x,x^{\prime },s)ds,
\label{w68} \\
S(x,x^{\prime }) &=&(\gamma {\cal P}+m)\Delta (x,x^{\prime }),  \nonumber \\
\Delta (x,x^{\prime }) &=&\mbox{sign}(x_{0}-x_{0}^{\prime })\int_{\Gamma
}f(x,x^{\prime },s)ds.  \label{w69}
\end{eqnarray}
One can see that $f(x,x^{\prime },s)$ obeys the following equations, 
\begin{eqnarray}  \label{w70}
&&-i\frac{d}{ds}\,f(x,x^{\prime },s)=\left( {\cal P}^{2}-m^{2}-\frac{q}{2}%
\sigma ^{\mu \nu }{\cal F}_{\mu \nu }\right) f(x,x^{\prime },s)\quad ,
\label{w72} \\[0.3cm]
&&\lim_{s\rightarrow +0}f(x,x^{\prime },s)=i\delta ^{(d)}(x-x^{\prime
})\quad .
\end{eqnarray}
Thus $f(x,x^{\prime },s)$ is the Fock-Schwinger function \cite{Sch1,Fock}.
So, representation (\ref{w68}) of causal Green function has the well-known
Schwinger form \cite{Sch1}. The concrete representation (\ref{w69}) of the
commutation function has the same form as the general representation \cite
{GG1996b}. Similar to \cite{GG1996b} one can select all singularities of
this function and see that it is continuous at $x_{0}-x_{0}^{\prime }$. In $%
d=4$ case one can transform this representation to the Fock form \cite{Fock}%
. The analysis of space-time singularities in the integrals (\ref{w66}) over
the contour $\Gamma _{1}^{a}$ can be done in a similar manner to \cite
{GG1996b}. In the $d=4$ the representations (\ref{w65}) - (\ref{w69})
coincide with the ones found in \cite{GGS}. To find the proper time
representation we used the special behavior of solutions (\ref{w14}). But
one has the different form of solutions (\ref{w14b}) if $d$ is odd, $E=0$
and all the imaginary eigenvalues $H_{j}$ of the field tensor are not equal
to zero. In this case since solutions (\ref{w14b}) can represent a special
case of the $d+1$ general form (\ref{w14}), one does not need to calculate
Green functions of the case independently and representations (\ref{w65}), (%
\ref{w68}) and (\ref{w69}) are valid if one puts $E=0$, $f((nx))=0$ and
replaces $qE/\sinh (qEs)\rightarrow qH_{(d-1)/2}/\sin \left(
qH_{(d-1)/2}\right) $ in formulas (\ref{w63}) and (\ref{w64}) (the functions 
$S^{a,p}(x,x^{\prime })$ are equal to zero).

\section{Some physical applications}

All the information about the processes of particle creation, annihilation,
and scattering in an external field (without radiative corrections) can be
extracted from the matrices $G\left( {}_{\zeta }|{}^{\zeta ^{\prime
}}\right) $ (\ref{we10}). These matrices define a canonical transformation
between in and out creation and annihilation operators in the generalized
Furry representation \cite{GG1977,FGS}, 
\begin{eqnarray}
&&a^{\dagger }(out)=a^{\dagger }(in)G\left( {}_{+}|{}^{+}\right)
+b(in)G\left( {}_{-}|{}^{+}\right) ,  \nonumber \\
&&b(out)=a^{\dagger }(in)G\left( {}_{+}|{}^{-}\right) +b(in)G\left(
{}_{-}|{}^{-}\right) .  \label{we10b}
\end{eqnarray}
Here $a_{\{n\}}^{\dagger }(in)$, $b_{\{n\}}^{\dagger }(in)$, $a_{\{n\}}(in)$%
, $b_{\{n\}}(in)$ are creation and annihilation operators of in-particles
and antiparticles respectively and $a_{\{n\}}^{\dagger }(out)$,$%
b_{\{n\}}^{\dagger }(out)$, $a_{\{n\}}(out),\;b_{\{n\}}(out)$ are ones of
out-particles and antiparticles, ${\{n\}}$ are possible quantum numbers. For
example, let us calculate the mean numbers of antiparticles created (which
are also equal to the numbers of pairs created) by the external field from
the in-vacuum $|0,in>$ with a given quantum number $p_{D},n,r$. By using
relations (\ref{we10b}) and (\ref{w18}) one finds representation of this
quantity: 
\begin{eqnarray}
&&N_{p_{D},n,r}=<0,in|b_{p_{D},n,r}^{\dagger }(out)b_{p_{D},n,r}(out)|0,in>=
\nonumber \\
&&\left. \left( G\left( {}_{+}|{}^{-}\right) ^{\dagger }G\left(
{}_{+}|{}^{-}\right) \right) _{p_{D},n,r,p_{D}^{\prime },n^{\prime
},r}\right| _{p_{D}=p_{D}^{\prime },\;n=n^{\prime }}\;,  \label{w73} \\
&&\left( G\left( {}_{+}|{}^{-}\right) ^{\dagger }G\left(
{}_{+}|{}^{-}\right) \right) _{p_{D},n,r,p_{D}^{\prime },n^{\prime },r}= 
\nonumber \\
&&\int_{-\infty }^{+\infty }\left( G\left( {}_{+}|{}^{-}\right) G\left(
{}_{+}|{}^{-}\right) ^{\dagger }\right) _{p_{-},n,r,p_{-},n^{\prime
},r}M^{\ast }(p_{D},p_{-})M(p_{D}^{\prime },p_{-})\frac{dp_{-}}{2\pi qE}.
\label{w74}
\end{eqnarray}
where the standard space coordinate volume regularization was used, so that $%
\delta (p_{j}-p_{j}^{\prime })\rightarrow \delta _{p_{j},p_{j}^{\prime }}$.
Then, by using formulas (\ref{w22c}) and (\ref{w22d}) one gets 
\begin{equation}
N_{p_{D},n,r}=\left. \int_{-\infty }^{+\infty }{}^{-}{\cal D}_{nn}M^{\ast
}(p_{D},p_{-})M(p_{D}^{\prime },p_{-})\frac{dp_{-}}{2\pi qE}\right|
_{p_{D}=p_{D}^{\prime }}\;.  \label{w75}
\end{equation}
If $L_{D}$ is the length of the correspondent edge of the space box then the
maximum wave length of the plane wave is $2L_{D}$. Thus, using the
Fourier-series expansion of ${}^{-}{\cal D}$ one gets 
\begin{equation}
N_{p_{D},n,r}=\frac{1}{L_{D}}\int_{0}^{L_{D}}{}^{-}{\cal D}_{nn}dp_{-}.
\label{w76}
\end{equation}
One can calculate the quantities ${}_{+}^{-}{\cal D}_{nn^{\prime }}$, given
by Eq. (\ref{w22d}), taking into account that the operator $\pi _{\perp }$
is Hermitian, and 
\[
e^{iK^{\ast }\pi _{\bot }}e^{-iK\pi _{\bot }}=\exp \left\{ -\frac{i}{2}%
K^{\ast }qFK-i\left( K-K^{\ast }\right) \pi _{\bot }\right\} . 
\]
Then all the integrals over $x^{j}$ in form (\ref{w22d}) can be expressed in
terms of the Laguerre polynomials \cite{GR}. In the special case when $%
n=n^{\prime }$ one has 
\begin{eqnarray}
&&{}^{-}{\cal D}_{nn}=\exp \left\{ -\pi \lambda -\mbox{Im}%
\;{}^{-}K(x_{-})\left( qf(p_{-}/qE)+qF\mbox{Re}\;{}^{-}K(x_{-})\right)
-\sum_{j=1}^{[d/2]-1}h_{j}\right\} \prod_{j=1}^{[d/2]-1}L_{n_{j}}(2h_{j}),
\label{w77} \\
&&h_{j}=-|qH_{j}|((\mbox{Im}\;{}^{-}K_{2j-1}(x_{-}))^{2}+(\mbox{Im}%
\;{}^{-}K_{2j}(x_{-}))^{2}),\;\;\pi _{-}>0.  \nonumber
\end{eqnarray}
Remember that we are discussing the case in which the constant electric
field acts for an infinite time. However, one can analyze the problem in
finite times $T=x_{out}^{0}-x_{in}^{0}$, acting similar to \cite{GG1996a}.
Then the mean numbers of pairs created by the external field $N_{p_{D},n,r}$
are the same if time $T$ is large enough: $\sqrt{qE}\,T\gg 1,\;\sqrt{qE}T\gg
\lambda \;\mbox{and}\;qET\gg |p_{D}|.$

Summing over the quantum numbers in (\ref{w76}), one can find the total
number of pairs created from the vacuum. Using standard regularization with
respect to the $(d-1)$-dimensional spatial volume $V_{(d-1)}$ and special
regularization with respect to time $T$ of acting of a constant electric
field \cite{GG1996a}, where $\int dp_{D}N_{p_{D},n,r}=qETN_{p_{D},n,r}$, one
gets 
\begin{eqnarray}
&&N=V_{(d-1)}n^{cr},  \nonumber \\
&&n^{cr}=J_{(d)}\frac{Tm^{2}\beta (1)}{2^{(d-1)}\pi ^{d/2}}\frac{E}{E_{c}}%
\exp \left\{ -\pi \frac{E_{c}}{E}\right\} ,
\end{eqnarray}
where $n^{cr}$ is the number density of the created pairs for time $T$, and
the coefficient $\beta (1)$ is defined by the next formula as a special case
on $l=1$: 
\begin{eqnarray}
&&\beta (l)=\prod_{j=1}^{(d-2)/2}\left\{ qH_{j}\coth (l\pi H_{j}/E)\right\}
\;,\;\;d\;{\rm is\;even}\;,  \nonumber \\
&&\beta (l)=\left( \frac{m^{2}E}{n\pi E_{c}}\right) ^{\frac{1}{2}%
}\;\prod_{j=1}^{(d-3)/2}\left\{ qH_{j}\coth (l\pi H_{j}/E)\right\} \;,\;\;d\;%
{\rm is\;odd}\;.  \label{w79}
\end{eqnarray}
Here, $E_{c}=m^{2}/|q|$ is the characteristic value of a constant electric
field strength. This quantity $N$ does not depend on the parameters of
plane-wave field and is the same as the number of pairs created in a
constant and uniform field \cite{GG1996a}. The corresponding formulas for
the $d=4$ case were first written in \cite{Nik}, and, in fact, can be
derived easily from the Schwinger formulas \cite{Sch1}.

By using proper-time kernel $f(x,x^{\prime },s)$ (\ref{w63}) one can
construct the $d$ dimensional form of the Schwinger out-in effective action 
\cite{Sch1} 
\begin{equation}
\Gamma _{out-in}=\frac{1}{2}\mbox{tr}\left\{ \int dx\int_{0}^{\infty
}s^{-1}\,f(x,x,s)\,ds\right\} \;.  \label{w80}
\end{equation}
It is not dependent on the parameters of the plane-wave field. And it is not
amazing because the effective action is a function of the field invariants
only which do not depend on the plane wave. Taking into account formulas (%
\ref{w55}) one can find the trace in (\ref{w80}) representation 
\begin{equation}
\rho (s)=\mbox{tr}\left\{ \exp \left( -i\frac{q}{2}\sigma ^{\mu \nu }F_{\mu
\nu }s\right) \right\} =2^{[d/2]}\cosh (qEs){\prod_{j=1}^{[(d-2)/2]}\limits}%
\cos (qH_{j}s).  \label{w81}
\end{equation}
Then, one calculates the probability for a vacuum to remain a vacuum by
using Schwinger method \cite{Sch1}, 
\begin{eqnarray}
&&P_{v}=\exp \left( -2\mbox{Im}\Gamma _{out-in}\right) =\exp \left\{ -\mu
N\right\} ,\;\;  \label{w82} \\
&&\mu =\sum_{l=0}^{\infty }\frac{\beta (l+1)}{(l+1)\beta (1)}\exp \left\{
-l\pi \frac{E_{c}}{E}\right\} \;,  \nonumber
\end{eqnarray}
where $\beta (l)$ is defined by (\ref{w79}). This result coincides with the
result from \cite{GG1996a}, and in $d=4$ the results coincide with Refs. 
\cite{Sch1,Nik}.

Let the operator of the current of the Dirac field operator $\psi (x)$ have
the form 
\begin{equation}
j_{\mu }=\frac{q}{2}\left[ \bar{\psi}(x),\gamma _{\mu }\;\psi (x)\right] ,
\label{w83}
\end{equation}
and operator of metric energy-momentum tensor (EMT) of the Dirac field
operator has the form 
\begin{eqnarray}
&&T_{\mu \nu }=\frac{1}{2}\left( T_{\mu \nu }^{can}+T_{\nu \mu
}^{can}\right) ,\;,  \label{w84} \\
&&T_{\mu \nu }^{can}=\frac{1}{4}\left\{ \left[ \bar{\psi}(x),\gamma _{\mu }%
{\cal P}_{\nu }\;\psi (x)\right] +\left[ {\cal P}_{\nu }^{\ast }\bar{\psi}%
(x),\gamma _{\mu }\;\psi (x)\right] \right\} \;,  \nonumber
\end{eqnarray}
where $T_{\mu \nu }^{can}$ is the canonical EMT operator. We are going to
discuss the following matrix elements with these operators: 
\begin{eqnarray}
&<&j_{\mu }>^{c}=<0,out|j_{\mu }|0,in>c_{v}^{-1}\;,  \label{w3.82} \\
&<&T_{\mu \nu }>^{c}=<0,out|T_{\mu \nu }|0,in>c_{v}^{-1}\;,  \label{w3.83} \\
&<&j_{\mu }>^{in}=<0,in|j_{\mu }|0,in>\;,  \label{w3.84} \\
&<&T_{\mu \nu }>^{in}=<0,in|T_{\mu \nu }|0,in>\;,  \label{w3.85} \\
&<&j_{\mu }>^{out}=<0,out|j_{\mu }|0,out>\;,  \label{w3.86} \\
&<&T_{\mu \nu }>^{out}=<0,out|T_{\mu \nu }|0,out>\;.  \label{w3.87}
\end{eqnarray}
Using the Green functions which were found before, one can present these
matrix elements in the following form: 
\begin{eqnarray}
&<&j_{\mu }>^{c}=iq\mbox{tr}\left\{ \gamma _{\mu }S^{c}(x,x)\right\}
=iq\left. \mbox{tr}\left\{ \gamma _{\mu }\gamma ^{\kappa }{\cal P}_{\kappa
}\Delta ^{c}(x,x^{\prime })\right\} \right| _{x=x^{\prime }}\;,
\label{w3.88} \\
&<&T_{\mu \nu }>^{c}=i/4\left. \mbox{tr}\left\{ \left( \gamma _{\mu }\left( 
{\cal P}_{\nu }+{\cal P^{\prime }}_{\nu }^{\ast }\right) +\gamma _{\nu
}\left( {\cal P}_{\mu }+{\cal P^{\prime }}_{\mu }^{\ast }\right) \right)
S^{c}(x,x^{\prime })\right\} \right| _{x=x^{\prime }}  \nonumber \\
&=&i\left. \mbox{tr}\left\{ B_{\mu \nu }\Delta ^{c}(x,x^{\prime })\right\}
\right| _{x=x^{\prime }}\;,  \label{w3.89} \\
&<&j_{\mu }>^{in}=<j_{\mu }>^{c}-<j_{\mu }>^{a}\;,\;\;\;  \label{w3.90a} \\
&<&j_{\mu }>^{out}=<j_{\mu }>^{c}-<j_{\mu }>^{p}\;,  \label{3.90b} \\
&<&T_{\mu \nu }>^{in}=<T_{\mu \nu }>^{c}-<T_{\mu \nu }>^{a}\;,\;\;\;
\label{w3.91a} \\
&<&T_{\mu \nu }>^{out}=<T_{\mu \nu }>^{c}-<T_{\mu \nu }>^{p}\;,
\label{w3.91b} \\
&<&j_{\mu }>^{(a,p)}=iq\left. \mbox{tr}\left\{ \gamma _{\mu }\gamma ^{\kappa
}{\cal P}_{\kappa }\Delta ^{a,p}(x,x^{\prime })\right\} \right|
_{x=x^{\prime }}\;,  \label{w3.92} \\
&<&T_{\mu \nu }>^{(a,p)}=i\left. \mbox{tr}\left\{ B_{\mu \nu }\Delta
^{a,p}(x,x^{\prime })\right\} \right| _{x=x^{\prime }}\;,  \label{w3.93} \\
&&B_{\mu \nu }=1/4\left\{ \gamma _{\mu }\left( {\cal P}_{\nu }+{\cal %
P^{\prime }}_{\nu }^{\ast }\right) +\gamma _{\nu }\left( {\cal P}_{\mu }+%
{\cal P^{\prime }}_{\mu }^{\ast }\right) \right\} \gamma ^{\kappa }{\cal P}%
_{\kappa }\;,  \nonumber
\end{eqnarray}
where the Green functions are given by Eqs. (\ref{w66}), (\ref{w68}) and the
relation 
\[
\Delta ^{c}(x,x)=\frac{1}{2}\left[ \Delta ^{-}(x,x)-\Delta ^{+}(x,x)\right] 
\]
is used. It is convenient to represent $<j_{\mu }>^{a,p}$ and $<T_{\mu \nu
}>^{a,p}$ as follows: 
\begin{eqnarray}
&&-<j_{\mu }>^{a}=<j_{\mu }>^{(1)}+<j_{\mu }>^{(2)}\;,\;\;\;  \label{w3.90aa}
\\
&&-<j_{\mu }>^{p}=<j_{\mu }>^{(1)}-<j_{\mu }>^{(2)}\;,  \label{3.90ba} \\
&&-<T_{\mu \nu }>^{a}=<T_{\mu \nu }>^{(1)}+<T_{\mu \nu }>^{(2)}\;,\;\;\;
\label{w3.91aa} \\
&&-<T_{\mu \nu }>^{p}=<T_{\mu \nu }>^{(1)}-<T_{\mu \nu }>^{(2)}\;,
\label{w3.91ba}
\end{eqnarray}
where 
\begin{eqnarray}
&<&j_{\mu }>^{(1)}=iq\left. \mbox{tr}\left\{ \gamma _{\mu }\gamma ^{\kappa }%
{\cal P}_{\kappa }\Delta ^{(1)}(x,x^{\prime })\right\} \right| _{x=x^{\prime
}}\;,  \label{w3.92a} \\
&<&T_{\mu \nu }>^{(1)}=i\left. \mbox{tr}\left\{ B_{\mu \nu }\Delta
^{(1)}(x,x^{\prime })\right\} \right| _{x=x^{\prime }}\;,  \label{w3.93a} \\
&&\Delta ^{(1)}(x,x^{\prime })=-\frac{1}{2}\,\int_{\Gamma _{3}+\Gamma
_{2}+\Gamma _{a}}\,f(x,x^{\prime },s)ds\quad ,  \label{wa42}
\end{eqnarray}
and all contributions with derivatives of $\Theta (\pm y^{D})$ functions,
which are formally divergent, are included in terms $<j_{\mu }>^{(2)}$ and $%
<T_{\mu \nu }>^{(2)}$. The nature of such divergences is connected with
infinite time $T$ of acting of a constant electric field and was discussed
in \cite{GG1996a} (see also below).

The components $<j_{\mu }>^{(1,2)}$ and $<T_{\mu \nu }>^{(1,2)}$ can not be
calculated in the framework of the perturbation theory with respect to the
external background or in the framework of the WKB method. Among them only
the term $<j_{\mu }>^{in}$ was calculated before in $d=4$ \cite{FGS}. Only
expression (\ref{w3.89}) for $<T_{\mu \nu }>^{c}$ has to be regularized and
renormalized because of the ultraviolet divergences. Expression (\ref{w3.88}%
) for the term $<j_{\mu }>^{c}$ is finite after the regularization lifting
and equal to zero. The terms $<j_{\mu }>^{(1)}$ and $<T_{\mu \nu }>^{(1)}$
are also finite, and $<j_{\mu }>^{(1)}=0.$ The terms $<j_{\mu }>^{(2)}$ and $%
<T_{\mu \nu }>^{(2)}$ have to be regularized with respect to time $T$ of
acting of a constant electric field \cite{GG1996a} and do not have standard
ultraviolet divergences. That is consistent with the fact that the
ultraviolet divergences have a local nature and result (as in the theory
without external field) from the leading local terms at $s\rightarrow +0$.
The nonzero contributions to the expressions $<j_{\mu }>^{(2)}$ and $<T_{\mu
\nu }>^{(1,2)}$ are related to global features of the theory and indicate
the vacuum instability.

At asymptotic region $x_{0}=T/2\rightarrow +\infty $ the densities of
current and the EMT of particles created are 
\begin{eqnarray}
&&j_{\mu }^{cr}=\frac{\int d{\bf x}\left( <j_{\mu }>^{in}-<j_{\mu
}>^{out}\right) }{\int d{\bf x}}\;,\;\;  \label{w3.97} \\
&&T_{\mu \nu }^{cr}=\frac{\int d{\bf x}\left( <T_{\mu \nu }>^{in}-<T_{\mu
\nu }>^{out}\right) }{\int d{\bf x}}\;,\;  \label{w3.98}
\end{eqnarray}
according to definitions (\ref{w3.84}) - (\ref{w3.87}).

Then, using representations (\ref{w3.90a}) - (\ref{w3.93}), and taking into
account terms $<j_{\mu }>^{(2)}$ and $<T_{\mu \nu }>^{(2)}$ which are
uniform, one gets from (\ref{w3.97}) and (\ref{w3.98}), 
\begin{eqnarray}
&&j_{\mu }^{cr}=2<j_{\mu }>^{(2)}\;,\;  \label{w3.99} \\
&&T_{\mu \nu }^{cr}=2<T_{\mu \nu }>^{(2)}\;.  \label{w3.100}
\end{eqnarray}
Using special regularization with respect to time $T$ of acting of a
constant electric field \cite{GG1996a} we can interpret these divergent
terms and, correct to first order of $\sqrt{qE}T$ , obtain 
\begin{eqnarray}
&<&j_{\mu }>^{cr}=2|q|n^{cr}\delta _{\mu }^{D},  \label{w86} \\
&<&T_{00}>^{cr}=<T_{DD}>^{cr}=qETn^{cr}\;.  \label{w87}
\end{eqnarray}
Other components of $<T_{\mu \nu }>^{cr}$ are of the order of $\ln (\sqrt{qE}%
T)$.

To study the backreaction of particles created on the electromagnetic field
and metrics one needs the expressions $<j_{\mu }>^{in}$ and $<T_{\mu \nu
}>^{in}.$ We found the form of $<j_{\mu }>^{in}=<j_{\mu }>^{(2)}=1/2<j_{\mu
}>^{cr}$ and $<T_{\mu \nu }>^{(2)}=1/2<T_{\mu \nu }>^{cr}.$ Let us calculate
other terms in expressions (\ref{w3.91a}). By using formulas (\ref{w55}) one
can find traces in (\ref{w3.89}) and (\ref{w3.93}) representations. One has
the next nonzero traces: one presented by expression (\ref{w81}) and 
\begin{eqnarray}
&&\mbox{tr}\left\{ \gamma ^{0}\gamma ^{D}\exp \left( -i\frac{q}{2}\sigma
^{\mu \nu }F_{\mu \nu }s\right) \right\} =\tanh (qEs)\rho (s)\;,  \nonumber
\label{w88} \\
&&\mbox{tr}\left\{ \gamma ^{2j}\gamma ^{2j-1}\exp \left( -i\frac{q}{2}\sigma
^{\mu \nu }F_{\mu \nu }s\right) \right\} =-\tan (qH_{j}s)\rho (s)\;.
\end{eqnarray}
Then, nonzero components of $<T_{\mu \nu }>^{c}$ and $<T_{\mu \nu }>^{(1)}$
are 
\begin{eqnarray}
&<&T_{\mu \nu }>^{c}=\int_{\Gamma _{c}}\tau _{\mu \nu }(s)ds\;, \\
&<&T_{\mu \nu }>^{(1)}=-\frac{1}{2}\int_{\Gamma _{3}+\Gamma _{2}+\Gamma
_{a}}\tau _{\mu \nu }(s)ds\;, \\
&&\tau _{\mu \nu }(s)=b_{\mu }(s)\rho (s)f^{(0)}(x,x,s)\;\mbox{if}\;\mu =\nu
\;,  \nonumber \\
&&b_{D}(s)=-b_{0}(s)=\frac{qE}{\sinh (2qEs)},\;\;b_{2j}(s)=b_{2j-1}(s)=\frac{%
qH_{j}}{\sin (2qH_{j}s)}\;.  \nonumber
\end{eqnarray}

\section{Acknowledgments}

D.M.G. thanks the Brazilian foundations CNPq for support. S.P.G. thanks the
Brazilian foundation CAPES for support, and the Department of Physics of UEL
and Department of Mathematical Physics of USP for their hospitality.

\end{document}